%% file: main.tex
\documentclass{jfm}
\usepackage{graphicx}
\usepackage{hyperref}
\usepackage{epstopdf, epsfig}
\usepackage{natbib}
\usepackage{tabularx}
\usepackage{subcaption}
\usepackage{amsmath}
\usepackage{bbm}
\usepackage{bm}
\usepackage[normalem]{ulem}
\usepackage[english]{babel}
\usepackage{tikz}
\usepackage{xcolor}
\colorlet{draft}{black}
\colorlet{revision1}{black}
\colorlet{revision2}{black}
\colorlet{revision3}{black}
\definecolor{DNS}{HTML}{000000}
\definecolor{df050}{HTML}{450A69}
\definecolor{df100}{HTML}{71196E}
\definecolor{df150}{HTML}{9B2964}
\definecolor{df200}{HTML}{C43C4E}
\definecolor{df250}{HTML}{E55C30}
\definecolor{df300}{HTML}{F8870E}
\definecolor{var1}{HTML}{4435A0}
\definecolor{var2}{HTML}{92558B}
\definecolor{var3}{HTML}{E07163}
\definecolor{var4}{HTML}{F8870E}

\DeclareRobustCommand\stl{\tikz[baseline]\draw[solid] (0,.5ex)--++(5ex,0) ;}

\DeclareRobustCommand\circles{\tikz \draw (0.5,0.1) circle (3pt);}
\DeclareRobustCommand\squares{\tikz \draw (0,0) rectangle (0.2,0.2);}
\DeclareRobustCommand\diamonds{\tikz \draw[rotate=45] (0,0) rectangle (0.2,0.2);}
\DeclareRobustCommand\triangles{\tikz \draw (0,0) -- (0.1,0.2) -- (0.2,0) -- cycle;}

\DeclareRobustCommand\stlcircles{\tikz[baseline]\draw[solid] (0,.5ex)--++(5ex,0) node[pos=0.5] {\circles} ;}
\DeclareRobustCommand\stlsquares{\tikz[baseline]\draw[solid] (0,.5ex)--++(5ex,0) node[pos=0.5] {\squares} ;}
\DeclareRobustCommand\stldiamonds{\tikz[baseline]\draw[solid] (0,.5ex)--++(5ex,0) node[pos=0.5] {\diamonds} ;}
\DeclareRobustCommand\stltriangles{\tikz[baseline]\draw[solid] (0,.5ex)--++(5ex,0) node[pos=0.5] {\triangles} ;}

\newcommand{\kernel}[0]{\mathcal{G}}

\title{Modeling the wall-slip in large eddy simulations with immersed boundaries}
\author{
    Morrison Rickard\aff{1}
    \and
    M. Houssem Kasbaoui\aff{1}\corresp{\email{houssem.kasbaoui@asu.edu}}}
\shortauthor{M. Rickard, and M. H. Kasbaoui}

\affiliation{
  \aff{1} School for Engineering of Matter, Transport and Energy, Arizona State University, Tempe, AZ 85281, USA.
}

\begin{document}

\maketitle

\begin{abstract}
	We present a framework for Large Eddy Simulations (LES) with Immersed Boundaries (IBs) to simulate high Reynolds number flows over complex walls. In this approach, which we call Immersed Boundary-Modeled LES (IBMLES), we volume-filter the Navier-Stokes equations to derive the IB bodyforce. This also gives rise to the subfilter-scale (SFS) stress and residual viscous stress tensors, the latter of which is closed, and we expect the SFS stress can be closed with existing models. We show that the IB bodyforce can be closed by modeling the wall-slip velocity and provide two such models. The first is an algebraic model based on volume-filtering the Van-Driest velocity profile, and the second is a slip-length model. We perform an \textit{a priori} analysis by volume-filtering direct numerical simulation (DNS) data of turbulent channel flow at $\Rey_{\tau}=5200$ to inform these models and investigate the behavior of the other terms in this formulation. We find streamwise wall-slip velocity is significant and both models show good agreement with volume-filtered DNS data on average. Slip velocity is non-uniform and retains a signature of inner or large scale flow structures depending on filter size. SFS terms are analogous to those in traditional LES and can likely be modeled with existing SFS models such Dynamic-Smagorinsky. Residual viscous stress is significant, so it must be considered in IBMLES. We perform filtering with multiple filter types and find there is little sensitivity to the choice of filter kernel so long as it abides by the assumptions given in this framework. Based on these results, we propose a dynamic procedure for determining slip-length that will be assessed in future studies.
\end{abstract}


{\bf MSC Codes }  {\it(Optional)} Please enter your MSC Codes here

\input{introduction.tex}

\input{framework.tex}
\input{apriori.tex}
\input{models.tex}
\input{models_algebraic.tex}
\input{models_slip_length.tex}
\input{filter_type.tex}
\input{dynamic_slip.tex}
\input{conclusion.tex}

\noindent\textbf{Acknowledgments.} The authors acknowledge support from the US National Science Foundation CAREER award \#2442871 (CBET-FD).

\noindent\textbf{Declaration of Interests.} The authors report no conflict of interest.

\appendix
\input{appendix.tex}

\bibliography{references,references_houssem2}

\end{document}

%% file: introduction.tex
\section{Introduction}

Practical engineering and environmental flows often involve highly turbulent flows around complex walls. Predicting the dynamics of these flows in Direct Numerical Simulations (DNS) is very challenging as high Reynolds number ($\Rey$) wall-bounded flows require extremely high near-wall resolution to accurately resolve the boundary layer. Complex walls further compound the computational cost due to the need for computationally intensive procedures to build and manage the discretization mesh. To address these bottlenecks, we introduce in this paper a new framework that leverages synergistic turbulence closures and wall-modeling on immersed boundaries to reduce the computational cost of high $\Rey$ simulations with complex walls.

Despite the large expansion in computing power, DNS of many engineering flows remains out of reach. \citet{choiGridpointRequirementsLarge2012} estimate that the minimum number of grid points required to resolve all flow features of a turbulent boundary layer scales as $\Rey^{37/14}$, which illustrates the computational expense associated with fully-resolved simulations. Despite this challenge, there remains a sustained effort to push the upper boundary of accessible Reynolds numbers in DNS, mainly for academic flow configurations. For example, the frontier friction Reynolds number ($\Rey_\tau$) in DNS of turbulent channel flow stands today at $\Rey_\tau\approx 12000$ \citep{pirozzoliStreamwiseVelocityVariance2024} to the best of our knowledge. This is up from $\Rey_\tau\approx 10000$ \citep{hoyasWallTurbulenceHigh2022} and 8000 \citep{yamamotoNumericalEvidenceLogarithmic2018} just a few years ago. Nonetheless, many engineering and environmental flows operate at even higher Reynolds numbers, making DNS intractable \citep{maniPerspectiveStateAerospace2023,gocLargeEddySimulation2021,boseWallModeledLargeEddySimulation2018,larssonLargeEddySimulation2016,piomelliLargeEddySimulations2014}.

Modeling the smallest scales of the flow in Large Eddy Simulations (LES) may reduce the computational cost substantially. In wall-resolved LES (WRLES), the mesh in bulk regions of the flow is coarser than in DNS, but it is refined to DNS resolution near walls. This way, the minimum number of grid points is estimated to scale as $\Rey^{13/7}$ \citep{choiGridpointRequirementsLarge2012}. Although an improvement compared to DNS, this scaling is still too costly for simulations of many practical problems \citep{maniPerspectiveStateAerospace2023}. For $\Rey>O(10^5)$, \citet{piomelliLargeEddySimulations2014} points out that more than $90\%$ of the grid points are required to resolve less than $10\%$ of the computational domain. In light of this, modeling the near wall flow, in so-called wall-modeled LES (WMLES) can further reduce the computational cost considerably. \citet{choiGridpointRequirementsLarge2012} estimate that the minimum number of grid points required in WMLES scales linearly with $\Rey$. While this estimated scaling is valid for zero pressure gradient turbulent boundary layers, and is thus a conservative estimate for relevant engineering and environmental flow configurations, it serves to illustrate the computational effectiveness of WMLES compared to DNS and WRLES. This makes it feasible to simulate flows at significantly higher $\Rey$ than DNS and WRLES (e.g. channel flow at $\Rey_\tau=2\times 10^7$ \citep{chungLargeeddySimulationWall2009}) and enables predictive simulations of many practical problems \citep{maniPerspectiveStateAerospace2023,lozano-duranPerformanceWallModeledBoundaryLayerConforming2022,gocLargeEddySimulation2021}. 

In configurations involving arbitrarily shaped and possibly moving geometries, there are additional challenges beyond turbulence resolution that must be overcome to accurately predict flow behavior. When the geometry does not align with Cartesian grids, body-fitted meshes are often used to resolve the geometry. However, the resulting curvilinear meshes are not guaranteed to be orthogonal and require complex solution algorithms that increase computational overhead considerably \citep{cristalloCombinedImmersedBoundary2006,verziccoImmersedBoundaryMethods2023}. Further, LES in these configurations may suffer from additional errors that arise from non-commutativity of filtering and differentiation operators on non-uniform body-conformal meshes \citep{vanellaEffectGridDiscontinuities2008,jordanLargeEddySimulationMethodology1999,ghosalBasicEquationsLarge1995}. In the case of moving geometry, body-fitted grids require remeshing between each time step which increases computational overhead even further \citep{verziccoImmersedBoundaryMethods2023}.

The computational bottlenecks associated with body-conformal meshes can be eliminated using immersed boundary (IB) methods to represent complex walls. In most of these methods, a bodyforce is formulated, often in an ad-hoc way that depends on the details of the discretization, such that it imposes the no-slip boundary condition on walls \citep{kasbaouiDirectNumericalSimulations2021,gozaAccurateComputationSurface2016,kempeImprovedImmersedBoundary2012,breugemSecondorderAccurateImmersed2012,pinelliImmersedboundaryMethodsGeneral2010,mittalVersatileSharpInterface2008,yangEmbeddedboundaryFormulationLargeeddy2006,kimImmersedBoundaryMethod2006,uhlmannImmersedBoundaryMethod2005,gilmanovGeneralReconstructionAlgorithm2003,peskinImmersedBoundaryMethod2002,udaykumarSharpInterfaceCartesian2001,kimImmersedBoundaryFiniteVolumeMethod2001,fadlunCombinedImmersedBoundaryFiniteDifference2000,mohd-yusofCombinedImmersedboundaryBspline1997,peskinFlowPatternsHeart1972}. Since wall-grid alignment is no longer an issue with these methods, fast and scalable Cartesian grid solvers can be used to integrate the governing equations \citep{capuanoCostVsAccuracy2023}. There is now ample evidence that IB methods employed in DNS provide accurate, numerically robust, and efficient predictions for a wide variety of flows, including those with moving geometries. We refer interested readers to the extensive reviews by \citet{mittalImmersedBoundaryMethods2005}, \citet{griffithImmersedMethodsFluid2020}, and \citet{verziccoImmersedBoundaryMethods2023}.

Despite their successes in DNS, using IB methods in LES has proven much more challenging. First, the ad-hoc formulation of most IB methods results in an IB bodyforce that depends on the local grid and discretization. Often, this forces the use of uniform grids, which is incompatible with WRLES. Using IB methods on non-uniform grids may introduce spurious stress oscillations \citep{seoSharpinterfaceImmersedBoundary2011,gozaAccurateComputationSurface2016,leeSourcesSpuriousForce2011} that may corrupt the computation of turbulent viscosity near walls and possibly the entire flow \citep{liaoSimulatingFlowsMoving2010,caiCouplingTurbulenceWall2021}. Due to this, \citet{verziccoImmersedBoundaryMethods2023} suggests that using IBs in LES may only be possible with uniform grids and wall models adapted to IBs.

To enable WMLES with IBs it is imperative to review the IB bodyforce formulation. As most IB methods are intended for use with existing DNS solvers, the IB bodyforce is often formulated to impose a no-slip boundary condition, which is not valid with large filter sizes typically used in WMLES \citep{sagautLargeEddySimulation2006,jimenezLargeEddySimulationsWhere2000,boseDynamicSlipBoundary2014}. Alternatively, the no-slip condition can be replaced by a constraint on the wall shear stress that derives from a wall model. But perhaps a more suited way for IB methods is to replace the no-slip boundary condition by a \emph{slip} condition. Recent work shows that imposing a slip boundary condition is in fact equivalent to imposing a wall shear stress \citep{boseDynamicSlipBoundary2014,boseWallModeledLargeEddySimulation2018,baeDynamicSlipWall2019}. 

Recently, we have introduced a new framework for simulations of flows with IBs that is robust and accurate in DNS, does not require uniform grids, and offers a natural extension to WMLES \citep{daveVolumefilteringImmersedBoundary2023,kasbaouiHighfidelityMethodologyParticleresolved2025,daveCharacterizationForcingSubfilter2025}. In our framework, the IB bodyforce derives from applying a volume-filter to the Navier-Stokes equations, rather than numerical heuristics. Volume-filtering differs from conventional LES filtering only by its treatment of the solid-fluid interface. While this leads to the emergence of the IB bodyforce, it also leads to a closure problem: the subfilter-scale (SFS) tensor, resulting from volume-filtering the convective term, and the IB bodyforce itself are both unclosed. 
In \citep{daveVolumefilteringImmersedBoundary2023}, we showed that the SFS tensor can be neglected and the IB bodyforce can be closed using the no-slip boundary conditions for sufficiently small filters. This yields the DNS limit of the volume-filtering IB method which we showed to yield stable and accurate predictions for a wide range of problems with fixed, forcibly moving, and freely moving geometries, including dense and neutrally-buoyant particles \citep{daveVolumefilteringImmersedBoundary2023,kasbaouiHighfidelityMethodologyParticleresolved2025}. For filter sizes that are large compared to near-wall flow structures, modeling the SFS tensor and IB bodyforce is critical. In \citep{daveCharacterizationForcingSubfilter2025}, we applied the volume-filtering immersed boundary method to a pure advection equation with an immersed cylinder and showed that modeling the resulting SFS term improves the solution drastically at large filter sizes.

In this study, we complete the extension of our framework introduced in \citep{daveVolumefilteringImmersedBoundary2023} to WMLES with immersed boundaries. This manuscript deals primarily with wall-slip modeling on IBs using turbulence theory and explicitly filtered DNS data to develop a framework for future work with IBMLES. \textit{A posteriori} analyses are left for a follow-up study.

The manuscript is organized as follows. In section \ref{sec:framework}, we present the volume-filtered conservation equations, differences with traditional frameworks for LES, discuss unclosed terms, and strategies to close them. We pay particular attention to closures of the IB bodyfroce and discuss another term that is unique to the volume-filtering framework known as the residual viscous stress tensor. In section \ref{sec:a_priori}, we compare the relative magnitude of closed and unclosed terms in the case of turbulent channel flow at friction Reynolds number $\Rey_{\tau}=5200$ using the DNS data of \citet{leeDirectNumericalSimulation2015}. We volume-filter the latter using coarse filter sizes relevant to WMLES and varying in the range 50 to 300 wall units. Importantly, we show that the wall-slip velocity is significant at these sizes and must be modeled appropriately. We propose two such models in section \ref{sec:models}. The first is algebraic in nature and is based on the Van Driest velocity profile \citep{vandriestTurbulentFlowWall1956}. The second is a slip-length model that extends earlier work by \citet{boseDynamicSlipBoundary2014}. Further, we comment on the effect of the filter kernel shape on statistics of wall slip and our proposed models in section \ref{sec:effect_filter}. We propose a dynamic procedure for determining slip-length based on the results of sections \ref{sec:a_priori}, \ref{sec:models}, and \ref{sec:effect_filter} in section \ref{sec:dynamic}. Finally, we give concluding remarks in section \ref{sec:conclusion}.

%% file: framework.tex
\section{The volume-filtering framework}\label{sec:framework}
\subsection{Overview}
\begin{figure}
    \centering
    \includegraphics[width=4in]{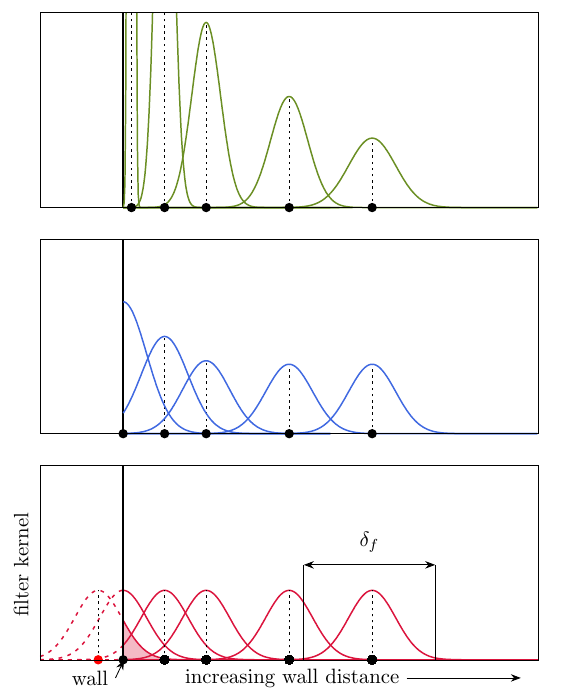}
    \caption{Schematic of filter kernel treatment near walls in WRLES, WMLES, and LES based on volume-filtering presented here (top to bottom). In WRLES, the filter is refined and rescaled to maintain unity with decreasing wall distance. In WMLES, the filter is truncated without changing its size and rescaled with decreasing wall distance. In the volume-filtering LES, the filter kernel is truncated without rescaling with decreasing wall distance. Filtered quantities are also extended a short distance into the solid (red point) so long as the kernel overlaps with the fluid region (shaded red region).}
    \label{fig:schematic_filters}
\end{figure}

Before presenting the mathematical details, it is worthwhile to start by emphasizing the main differences between different filtering methods and the connection with immersed boundary methods.

To help with this discussion, we show in figure \ref{fig:schematic_filters} the qualitative variation of the filter kernel with wall distance in three methods: WRLES (top panel), WMLES with the treatment of \citet{boseDynamicSlipBoundary2014} (middle panel), and the volume-filtering LES we present here (bottom panel). First, note that away from the wall, the filter shape and size, denoted by $\delta_f$, may be taken identical in all three methods. These methods differ only by their treatment of the wall region. In WRLES, the filter, be it either explicit or implicit, is refined to a DNS resolution close to walls. That is $\delta_f$ decreases approaching a wall, such that $\delta_f^+\lesssim 1$ within the boundary layer, where the `$+$' notation refers to non-dimensionalization with inner wall units using combinations of friction velocity $u_\tau$ and kinematic viscosity $\nu$. This ensures that near-wall turbulent flow structures are well resolved in WRLES. Shrinking the filter support closer to the wall requires rescaling the filter to ensure that it remains unitary, even though its shape is unchanged. In WMLES, the filter size is typically uniform such that $\delta_f^+=O(10^2)$ everywhere. Because of this, the filter may extend beyond the fluid region at distances less than $\delta_f/2$ from a wall. \citet{boseDynamicSlipBoundary2014} suggest truncating the portion of the filter that falls inside the solid, thereby altering the filter shape, and rescaling it to ensure that it remains unitary. The consequences of this treatment are discussed in detail in \citep{boseDynamicSlipBoundary2014}. 

In the volume-filtering LES we present here, we also maintain a uniform filter size, with a typical size $\delta_f^+=O(10^2)$, and truncate the part of the kernel that falls within the solid region. However, there are two essential differences from the approach of \citet{boseDynamicSlipBoundary2014}. First, we do not rescale truncated filters. Instead, we introduce a fluid volume fraction $\alpha_f$ to account for the deficit of mass and momentum near a wall. This quantity represents the fraction of the fluid region among the total region (solid and fluid) covered by the filter. For a planar wall, $\alpha_f=1/2$ at the wall since exactly half of the region covered by the filter is fluid. Away from walls, $\alpha_f=1$. Second, we extend our definitions of filtered quantities inside the solid region. The red point in the bottom panel of figure \ref{fig:schematic_filters} illustrates this aspect. The filtered velocity at this point represents the average velocity within the portion of the filter that falls in the fluid region (portion shaded in red). In general, filtered fluid quantities may be non-zero at points inside the solid but within $\delta_f/2$ from a wall.

Compared to other WMLES methods, volume-filtering LES has two significant advantages: (i) it eliminates commutation errors and (ii) turns boundary conditions, such as the no-slip condition, into volumetric forcing, effectively leading to an IB method. In contrast, WRLES and WMLES discussed above must contend with commutation errors. Since the filter size and/or shape vary with wall distance, filtering and derivative operators are no longer commutative \citep{germanoPhysicalEffectsVariable2002,kleinAnalysisModellingCommutation2020,sagautLargeEddySimulation2006,gullbrandGridindependentLargeeddySimulation2002}. This leads to additional terms in the filtered transport equations, which can be on the order of the subgrid terms
\citep{ghosalBasicEquationsLarge1995,jordanLargeEddySimulationMethodology1999,gullbrandGridindependentLargeeddySimulation2002,vanellaEffectGridDiscontinuities2008,kleinAnalysisModellingCommutation2020}. These commutation terms are ignored in most WRLES and WMLES methods. Volume-filtering and differentiation operations are not commutative either. The terms arising from the non commutative operations are part of what forms the IB bodyforce.

\subsection{Derivation}
We now present the mathematical details underpinning the volume-filtering framework, discuss the commutation terms, their connection with IB bodyforce, and the different closure problems. This treatment follows our earlier work in \citep{daveVolumefilteringImmersedBoundary2023,kasbaouiHighfidelityMethodologyParticleresolved2025}, where we first introduced the volume-filtering Immersed Boundary method. Unlike in \citep{daveVolumefilteringImmersedBoundary2023}, we make no assumptions on the relative size of the filter kernel $\delta_f$ with respect to the flow structures.

In the following, we assume that the fluid is Newtonian, has constant density $\rho_f$, and constant viscosity $\mu_f$. Thus, the equations of motion for this fluid are dictated by the incompressible Navier-Stokes equations,
\begin{gather}
    \nabla \cdot \boldsymbol{u}_f=0,\label{eq:NVS_1}\\
    \rho_f\left(\frac{\partial\boldsymbol{u}_f}{\partial t}+\nabla\cdot\left(\boldsymbol{u}_f \boldsymbol{u}_f\right)\right)=\nabla\cdot\bm{\mathsf{\tau}}_f, \label{eq:NVS_2}
\end{gather}
where $\boldsymbol{u}_f$ is the unfiltered fluid velocity, $\bm{\mathsf{\tau}}_f=-p \mathsfbi{I} + \mu_f\left(\nabla \boldsymbol{u}_f + \nabla \boldsymbol{u}_f^T\right)$ is the total stress tensor, and $p$ is the fluid pressure.  

To derive volume-filtered equations, we consider a filter kernel \(\kernel(\boldsymbol{y})\) for $\boldsymbol{y}\in\Re^3$ with size $\delta_f$. The filter kernel must satisfy two fundamental conditions
\begin{gather}
    \iiint_{\boldsymbol{y}\in\Re^3}\kernel(\boldsymbol{y})dV=1,\\
    \kernel(-\boldsymbol{y})=\kernel(\boldsymbol{y}).
\end{gather}
The first dictates that the filter is unitary while the second requires it to be symmetric. Although not strictly required for the derivation, it is also convenient to take a compact filter kernel, i.e., one that verifies the condition,
\begin{gather}
    \kernel(\boldsymbol{y})=0\quad\text{if}\:||\boldsymbol{y}||\geq\delta_f/2.
\end{gather}

We define the fluid volume fraction $\alpha_f$ and filtered fluid velocity \(\overline{\boldsymbol{u}}_f\) as follows
\begin{eqnarray}
    \alpha_f(\boldsymbol{x},t)&=&\iiint_{\boldsymbol{y}\in\Re^3}\mathbb{I}_f(\boldsymbol{y},t)\kernel(\boldsymbol{x}-\boldsymbol{y})dV,\label{eq:def_filtered_I}\\
    \alpha_f(\boldsymbol{x},t)\overline{\boldsymbol{u}}_f(\boldsymbol{x},t)&=&\iiint_{\boldsymbol{y}\in\Re^3}\mathbb{I}_f(\boldsymbol{y},t)\boldsymbol{u}_f(\boldsymbol{y},t)\kernel(\boldsymbol{x}-\boldsymbol{y})dV.\label{eq:def_filtered_u}
\end{eqnarray}
Here, \(\mathbb{I}_f(\boldsymbol{y},t)\) is the fluid-phase indicator function whose value is 1 if \(\boldsymbol{y}\) is in the fluid at time \(t\) and 0 otherwise. The function \(\mathbb{I}_f\) truncates the filtering region, as schematized in the bottom panel of figure \ref{fig:schematic_filters}. From equation (\ref{eq:def_filtered_I}), the volume fraction may be interpreted as the filtered indicator function. That is, \(\mathbb{I}_f\) is a sharp and discontinuous field, whereas $\alpha_f$ is a smooth field that diffuses the solid-fluid interface over a band of width $\delta_f$. Other fluid quantities may be filtered in the same manner as the fluid velocity in (\ref{eq:def_filtered_u}).

As previously discussed, filtering and differentiation operations are not commutative. This results from the truncation of the filtering region by the phase-indicator function \(\mathbb{I}_f\). Still, we can show that the following identities hold exactly \citep{daveVolumefilteringImmersedBoundary2023},
\begin{eqnarray}
    \alpha_f(\boldsymbol{x},t)\overline{\nabla\boldsymbol{\phi}}(\boldsymbol{x},t)&=&\nabla\left(\alpha_f\overline{\boldsymbol{\phi}}\right)-\iint_{\boldsymbol{y}\in S_w}\boldsymbol{n}\boldsymbol{\phi}(\boldsymbol{y},t)\kernel(\boldsymbol{x}-\boldsymbol{y})dS,\label{eq:identity_grad}\\
    \alpha_f(\boldsymbol{x},t)\overline{\nabla\cdot\boldsymbol{\phi}}(\boldsymbol{x},t)&=&\nabla\cdot\left(\alpha_f\overline{\boldsymbol{\phi}}\right)-\iint_{\boldsymbol{y}\in S_w}\boldsymbol{n}\cdot\boldsymbol{\phi}(\boldsymbol{y},t)\kernel(\boldsymbol{x}-\boldsymbol{y})dS,\label{eq:identity_div}\\
    \alpha_f(\boldsymbol{x},t)\overline{\frac{\partial\boldsymbol{\phi}}{\partial t}}(\boldsymbol{x},t)&=&\frac{\partial}{\partial t}\left(\alpha_f\overline{\boldsymbol{\phi}}\right)+\iint_{\boldsymbol{y}\in S_w}(\boldsymbol{n}\cdot\boldsymbol{u}_w)\boldsymbol{\phi}(\boldsymbol{y},t)\kernel(\boldsymbol{x}-\boldsymbol{y})dS,\label{eq:identity_dt}
\end{eqnarray}
where $\boldsymbol{\phi}(\boldsymbol{x})$ is a scalar or vector property of the fluid, $\boldsymbol{n}$ is a unitary vector normal to the wall and pointing from the solid to the fluid, $\boldsymbol{u}_w$ is the local wall velocity, and $S_w$ denotes all walls. It is also possible to rearrange the equations above to get exact expressions of the commutation terms
\begin{eqnarray}
    \overline{\nabla\boldsymbol{\phi}} - \nabla \overline{\boldsymbol{\phi}} &=& \frac{1}{\alpha_f}\left((\nabla \alpha_f) \overline{\boldsymbol{\phi}} -\iint_{\boldsymbol{y}\in S_w}\boldsymbol{n}\boldsymbol{\phi}(\boldsymbol{y},t)\kernel(\boldsymbol{x}-\boldsymbol{y})dS\right),\label{eq:identity_grad_ver2}\\
    \overline{\nabla\cdot\boldsymbol{\phi}}-\nabla\cdot\overline{\boldsymbol{\phi}} &=& \frac{1}{\alpha_f}\left((\nabla \alpha_f)\cdot \overline{\boldsymbol{\phi}}-\iint_{\boldsymbol{y}\in S_w}\boldsymbol{n}\cdot\boldsymbol{\phi}(\boldsymbol{y},t)\kernel(\boldsymbol{x}-\boldsymbol{y})dS\right),\label{eq:identity_div_ver2}\\
    \overline{\frac{\partial\boldsymbol{\phi}}{\partial t}}-\frac{\partial\overline{\boldsymbol{\phi}}}{\partial t} &=& \frac{1}{\alpha_f}\left( \left(\frac{\partial \alpha_f}{\partial t}\right) \overline{\boldsymbol{\phi}} +\iint_{\boldsymbol{y}\in S_w}(\boldsymbol{n}\cdot\boldsymbol{u}_w)\boldsymbol{\phi}(\boldsymbol{y},t)\kernel(\boldsymbol{x}-\boldsymbol{y})dS\right). \label{eq:identity_dt_ver2}
\end{eqnarray}

To obtain the volume-filtered transport equations, we apply the filtering operation to the transport equations (\ref{eq:NVS_1}) and (\ref{eq:NVS_2}). With help of  identities  (\ref{eq:identity_grad})--(\ref{eq:identity_dt}), we get,
\begin{eqnarray}
    \frac{\partial\alpha_f}{\partial t}+\nabla\cdot(\alpha_f\overline{\boldsymbol{u}}_f)&=&0,\\
    \rho_f\left(\frac{\partial}{\partial t}\left(\alpha_f\overline{\boldsymbol{u}}_f\right)+\nabla\cdot\left(\alpha_f\overline{\boldsymbol{u}_f\boldsymbol{u}_f}\right)\right)&=&\nabla\cdot\left(\alpha_f\overline{\bm{\mathsf{\tau}}}_f\right) \nonumber \\
    &&-\iint_{\boldsymbol{y}\in S_w}\boldsymbol{n}\cdot\bm{\mathsf{\tau}}_f(\boldsymbol{y},t)\kernel(\boldsymbol{x}-\boldsymbol{y})dS.\label{eq:filtered_momentum_ver1}
\end{eqnarray}
The filtered stress tensor $\overline{\bm{\mathsf{\tau}}}_f$  in the filtered momentum equation (\ref{eq:filtered_momentum_ver1}) requires special care. We decompose this tensor into two contributions, $\overline{\bm{\mathsf{\tau}}}_f=\overline{\bm{\mathsf{\tau}}}_f^R + \mathsfbi{R}_\mu$, where each of these reads
\begin{eqnarray}
    \overline{\bm{\mathsf{\tau}}}_f^R&=&-\overline{p} \mathsfbi{I} + \mu_f\left(\nabla \overline{\boldsymbol{u}}_f + \nabla \overline{\boldsymbol{u}}_f^T-\frac{2}{3}\left(\nabla\cdot\overline{\boldsymbol{u}}_f\right)\mathsfbi{I}\right),\\
    \mathsfbi{R}_\mu &=&\mu_f \left(\overline{\nabla\boldsymbol{u}_f+\nabla\boldsymbol{u}_f^T}\right) - \mu_f\left(\nabla \overline{\boldsymbol{u}}_f+\nabla \overline{\boldsymbol{u}}_f^T-(2/3)(\nabla \cdot\overline{\boldsymbol{u}}_f)\mathsfbi{I}\right).
\end{eqnarray}
The tensors $\overline{\bm{\mathsf{\tau}}}_f^R$ and \(\mathsfbi{R}_{\mu}\) as the so-called resolved fluid stress tensor and residual viscous stress tensor, respectively. While both tensors are closed, \(\mathsfbi{R}_{\mu}\) has a non-trivial form and is poorly understood. Using (\ref{eq:identity_grad_ver2}), (\ref{eq:identity_div_ver2}), and the no-slip condition $\boldsymbol{u}_f|_w=\boldsymbol{u}_w$, we get an alternative expression for \(\mathsfbi{R}_{\mu}\),
\begin{eqnarray}
    \mathsfbi{R}_\mu &=&\frac{\mu_f}{\alpha_f}\Biggl(
    (\nabla\alpha_f) \overline{\boldsymbol{u}}_f + \overline{\boldsymbol{u}}_f (\nabla\alpha_f) - (2/3)(\nabla\alpha\cdot \overline{\boldsymbol{u}}_f)\mathsfbi{I}\nonumber\\
    &&\hspace{8ex} 
    -\iint_{\boldsymbol{y}\in S_w} \Bigl(\boldsymbol{n}\boldsymbol{u}_w+\boldsymbol{u}_w\boldsymbol{n}-(2/3)(\boldsymbol{n}\cdot\boldsymbol{u}_w)\mathsfbi{I}\Bigr)\kernel(\boldsymbol{x}-\boldsymbol{y})dS
    \Biggr). \label{eq:residual_viscous_stress}
\end{eqnarray}
This symmetric tensor vanishes quickly away from the fluid-solid interface. Close to walls, \(\mathsfbi{R}_{\mu}\) scales with the difference between the filtered fluid velocity at the wall $\overline{\boldsymbol{u}}_f|_w$ and the wall velocity itself $\boldsymbol{u}_w$. We refer to this as the wall slip velocity: 
\begin{equation}
\boldsymbol{u}_\mathrm{slip}=\overline{\boldsymbol{u}}_f|_w- \boldsymbol{u}_w.
\end{equation}

As with other LES methods, the convective term \(\overline{\boldsymbol{u}_f\boldsymbol{u}_f}\) in  equation (\ref{eq:filtered_momentum_ver1}) presents a closure problem. Following standard LES strategies, we define the subfilter-scale stress tensor,
\begin{gather}
    \bm{\mathsf{\tau}}_{\mathrm{sfs}}=\overline{\boldsymbol{u}_f\boldsymbol{u}_f}-\overline{\boldsymbol{u}}_f\overline{\boldsymbol{u}}_f.
\end{gather}
Introducing this tensor in equation (\ref{eq:filtered_momentum_ver1}) and using the product rule on the first term on the right-hand side, we obtain the following form of the volume-filtered conservation equations
\begin{eqnarray}
\frac{\partial\alpha_f}{\partial t}+\nabla\cdot(\alpha_f\overline{\boldsymbol{u}}_f)&=&0,\label{eq:filtered_mass}\\
\rho_f\left(\frac{\partial}{\partial t}\left(\alpha_f\overline{\boldsymbol{u}}_f\right)+\nabla\cdot\left(\alpha_f\overline{\boldsymbol{u}}_f\overline{\boldsymbol{u}}_f\right)\right) &=& \alpha_f\nabla\cdot\left(\overline{\bm{\mathsf{\tau}}}^R_f+\mathsfbi{R}_{\mu}-\rho_f \bm{\mathsf{\tau}}_{\mathrm{sfs}}\right) +\boldsymbol{F}_\mathrm{IB} \label{eq:filtered_momenutm}
\end{eqnarray}
where $\boldsymbol{F}_\mathrm{IB}$ is a bodyforce that reads
\begin{equation}
\boldsymbol{F}_\mathrm{IB}(\boldsymbol{x},t) = -\iint_{\boldsymbol{y}\in S_w}\boldsymbol{n}\cdot\bm{\mathsf{\tau}}_f(\boldsymbol{y},t)\kernel(\boldsymbol{x}-\boldsymbol{y})dS
         +
         (\nabla\alpha_f)\cdot\left(\overline{\bm{\mathsf{\tau}}}^R_f+\mathsfbi{R}_{\mu}-\rho_f \bm{\mathsf{\tau}}_{\mathrm{sfs}}\right) \label{eq:forcing_ver1}
\end{equation}
Leveraging the identity in equation (\ref{eq:identity_grad}) with $\boldsymbol{\phi}=\boldsymbol{I}$, it is possible to approximate  $\boldsymbol{F}_\mathrm{IB}$  in the following manner
\begin{gather}
    \boldsymbol{F}_\mathrm{IB}(\boldsymbol{x},t)=-\iint_{\boldsymbol{y}\in S_w}\boldsymbol{n}\cdot\left(\bm{\mathsf{\tau}}_f-\overline{\bm{\mathsf{\tau}}}^R_f-\mathsfbi{R}_{\mu}+\rho_f \bm{\mathsf{\tau}}_\mathrm{sfs}\right)\kernel(\boldsymbol{x}-\boldsymbol{y})dS + O\left(\delta_f^2\right). \label{eq:forcing_ver2}
\end{gather}
This term represents a force density exerted by the immersed solid on the fluid. It is non-zero only in close vicinity to the fluid-solid interface. With a compact filter, this is exactly a band of size $\delta_f$ centered on the interface. The bodyforce $\boldsymbol{F}_\mathrm{IB}$ arises from stresses on the fluid-solid interface in this filtered framework. 

From a computational perspective, the term $\boldsymbol{F}_\mathrm{IB}$ may be thought of as the Immersed Boundary forcing since it serves to impose boundary conditions on the fluid-solid interface. As we have shown in \citep{daveVolumefilteringImmersedBoundary2023}, $\boldsymbol{F}_\mathrm{IB}$ imposes the no-slip boundary condition in the DNS limit, i.e., when $\delta_f\rightarrow 0$.

As anticipated, the filtered momentum equation (\ref{eq:filtered_momenutm}) requires closure. The unclosed terms here are (1) the subfilter-scale tensor $\bm{\mathsf{\tau}}_\mathrm{sfs}$ and (2) the IB bodyforce $\boldsymbol{F}_\mathrm{IB}$. The residual viscous stress tensor $\mathsfbi{R}_\mu$ is closed, albeit it has a non-trivial dependence on the filtered quantities that ought to be clarified. Since there have been numerous studies devoted to closing $\bm{\mathsf{\tau}}_\mathrm{sfs}$, the rest of this work focuses on $\boldsymbol{F}_\mathrm{IB}$ and $\mathsfbi{R}_\mu$, which represent terms unique to this method.

\subsection{Modeling the immersed boundary bodyforce}\label{sec:ib_modeling}
Closing the IB bodyforce $\boldsymbol{F}_\mathrm{IB}$ amounts to formulating a wall model since the wall stress $(\boldsymbol{n}\cdot\bm{\mathsf{\tau}}_f)|_w$ is the only unclosed term in expressions (\ref{eq:forcing_ver1}) and (\ref{eq:forcing_ver2}).

In principle, traditional wall models, such as algebraic wall models\textcolor{revision2}{, approximate boundary conditions \citep{piomelliNewApproximateBoundary1989,schumannSubgridScaleModel1975},} and thin boundary-layer equations \citep{boseWallModeledLargeEddySimulation2018}, could be adapted to this framework. These methods are typically formulated to predict the wall stress $(\boldsymbol{n}\cdot\bm{\mathsf{\tau}}_f)|_w$ from the filtered velocity $\overline{\boldsymbol{u}}_f$ and its gradients close to walls. To use these models in this framework, one would insert the wall stress $(\boldsymbol{n}\cdot\bm{\mathsf{\tau}}_f)|_w$ determined by these models in $\boldsymbol{F}_\mathrm{IB}$ via expression (\ref{eq:forcing_ver1}) or (\ref{eq:forcing_ver2}).

In this study, we propose an alternative method to wall modeling that is based on modeling the wall slip velocity $\boldsymbol{u}_\mathrm{slip}=\overline{\boldsymbol{u}}_f|_w-\boldsymbol{u}_w$, i.e., the difference between the filtered velocity at the solid-fluid interface and the actual wall velocity, rather than wall stress $(\boldsymbol{n}\cdot\bm{\mathsf{\tau}}_f)|_w$. Following the treatment in \citep{daveVolumefilteringImmersedBoundary2023}, we can show that the wall slip $\boldsymbol{u}_\mathrm{slip}$ and wall stress $(\boldsymbol{n}\cdot\bm{\mathsf{\tau}}_f)|_w$ are related to one another via
\begin{eqnarray}
  \left.(\boldsymbol{n}\cdot\bm{\mathsf{\tau}}_f)\right|_w &=& \left.\boldsymbol{n}\cdot
    \left(\overline{\bm{\mathsf{\tau}}}^R_f+\mathsfbi{R}_{\mu}-\rho_f \bm{\mathsf{\tau}}_\mathrm{sfs}\right)\right|_w \nonumber\\
    &&\hspace{-10ex}+
    \rho_f\frac{\alpha_{f,w}}{\Sigma_w}\left[\frac{d}{dt}\left(\boldsymbol{u}_w 
    +\boldsymbol{u}_\mathrm{slip}\right)-\left.\nabla\cdot\left(\overline{\bm{\mathsf{\tau}}}^R_f+\mathsfbi{R}_{\mu}-\bm{\mathsf{\tau}}_\mathrm{sfs}\right)\right|_w\right] +O(\delta_f^2),\label{eq:wall_stress}
\end{eqnarray}
where $\alpha_{f,w}$ and $\Sigma_{w}$ are the fluid volume fraction and surface density, respectively, at the wall. Here, the surface density function is
\begin{eqnarray}
  \Sigma(\boldsymbol{x},t) = \iint_{\boldsymbol{y}\in S_w}\kernel(\boldsymbol{x}-\boldsymbol{y})dS.\label{eq:SDF}
\end{eqnarray}
Integrating $\Sigma$ over the entire domain gives the total wall area.
For planar walls, $\alpha_{f,w}=1/2$ and $\Sigma_w=O(1/\delta_f)$, the precise value depending on the kernel $\kernel$.

In equation (\ref{eq:wall_stress}), the wall slip velocity $\boldsymbol{u}_\mathrm{slip}$ concentrates the unknown, as it is the only unclosed term. Thus, formulating a model for $\boldsymbol{u}_\mathrm{slip}$ is equivalent to specifying a wall model. This approach is better suited in immersed boundary methods, since most of these methods are formulated to impose a Dirichlet condition on the velocity via a volumetric forcing term.

It is important to realize that the no-slip boundary condition, i.e., $\boldsymbol{u}_\mathrm{slip}=0$, applies only when the filter size is very small. For wall bounded turbulence, this requires $\delta_f^+<1$, which we refer to as the DNS limit \citep{daveVolumefilteringImmersedBoundary2023,kasbaouiHighfidelityMethodologyParticleresolved2025}. 

\begin{figure}
    \centering
    \includegraphics[width=4in]{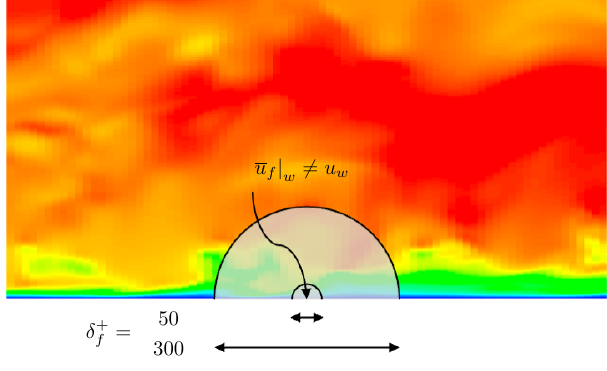}
    \caption{The filtered fluid velocity at the wall represents an average velocity of the flow in the region under the filter kernel centered at the wall (marked in gray). This leads to significant slip velocity at filter sizes $\delta_f^+=O(100)$.}
    \label{fig:schematic_wall_slip}
\end{figure}

For larger filter sizes, the no-slip condition ($\boldsymbol{u}_\mathrm{slip}=0$) is no longer valid. That is because the filtered velocity at the wall $\overline{\boldsymbol{u}}_f|_w$ represents the average fluid velocity in a volume of size $\delta_f$ centered at the wall. Figure 2 shows a schematic of the averaging volume and the flow structures that may be part of this region. For vanishingly small filter sizes ($\delta_f\rightarrow 0$), the averaging volume collapses onto a point which leads to $\overline{\boldsymbol{u}}_f|_w\rightarrow \boldsymbol{u}_w$. Conversely, increasing the filter size increases the slip as the averaging volume used to compute $\overline{\boldsymbol{u}}_f|_w$ covers increasingly more of the viscous, buffer, and logarithmic sublayers.

The residual viscous stress tensor and the slip velocity are also closely connected. Considering expression (\ref{eq:residual_viscous_stress}) at the wall, one can show
\begin{eqnarray}
    \left.{\mathsfbi{R}_\mu}\right|_w &=&\frac{\mu_f}{\alpha_{f,w}}
    \iint_{\boldsymbol{y}\in S_w} \Bigl(\boldsymbol{n}\boldsymbol{u}_\mathrm{slip}+\boldsymbol{u}_\mathrm{slip}\boldsymbol{n}-(2/3)(\boldsymbol{n}\cdot\boldsymbol{u}_\mathrm{slip})\boldsymbol{I}\Bigr)\kernel(\boldsymbol{x}-\boldsymbol{y})dS.\label{eq:residual_wall}
    \end{eqnarray}
Thus, $\mathsfbi{R}_\mu$ scales with the slip velocity. Since the latter is significant for filter sizes $\delta_f^+=O(100)$, $\mathsfbi{R}_\mu$ is also significant at those filter sizes and must be accounted for correctly.

%% file: apriori.tex
\section{Volume-filtered turbulent channel flow: \textit{a priori}} analysis\label{sec:a_priori}
In this section, we characterize the terms that arise in the volume-filtering framework described in \S\ref{sec:framework} using DNS data by \citet{leeDirectNumericalSimulation2015} of a turbulent channel flow at friction Reynolds number $\mathrm{Re}_{\tau}=u_\tau h/\nu=5200$, where $h$ is the channel half-height, $u_\tau=\sqrt{\tau_w/\rho_f}$ is the friction velocity, and $\tau_w$ is the wall shear stress. The static and planar walls in this case make it easier to analyze the volume-filtering terms without complication that may arise with complex and/or moving walls. We pay special attention to the effect of varying filter size $\delta_f$ on wall slip $\boldsymbol{u}_\mathrm{slip}=\overline{\boldsymbol{u}}_f|_w- \boldsymbol{u}_w$, interface forcing $\boldsymbol{F}_\mathrm{IB}$, and residual viscous stress tensor $\mathsfbi{R}_\mu$. We also characterize flow statistics and other terms appearing in the momentum equation (\ref{eq:filtered_momenutm}).

\subsection{Overview of the filtering procedure}

To filter the DNS data, we consider a kernel that expresses as the product of three one-dimensional kernels,
\begin{equation}
   \kernel(x,y,z)=\kernel_1(x)\kernel_1(y)\kernel_1(z).\label{eq:filter_decomposition}
\end{equation}
where $x$, $y$, and $z$ are the streamwise, wall-normal, and spanwise coordinates, respectively. Since the three-dimensional kernel $\kernel$ must be symmetric and unitary, the one-dimensional kernels $\kernel_1$ in (\ref{eq:filter_decomposition}) must also satisfy the same conditions. 

The data we present throughout this section results from filtering with the cosine kernel,
\begin{equation}
   \kernel_1(r)=\begin{cases} 
      \frac{\pi}{2\delta_f}\cos{\left(\frac{\pi r}{\delta_f}\right)} & |r|< \delta_f/2 \\
      0 & |r|\geq\delta_f/2 
   \end{cases}.\label{eq:cosine_filter}
\end{equation}
We also consider other kernels in \S\ref{sec:effect_filter} and show that there are no qualitative differences.

\begin{figure}\centering
  \includegraphics[width=5.2in]{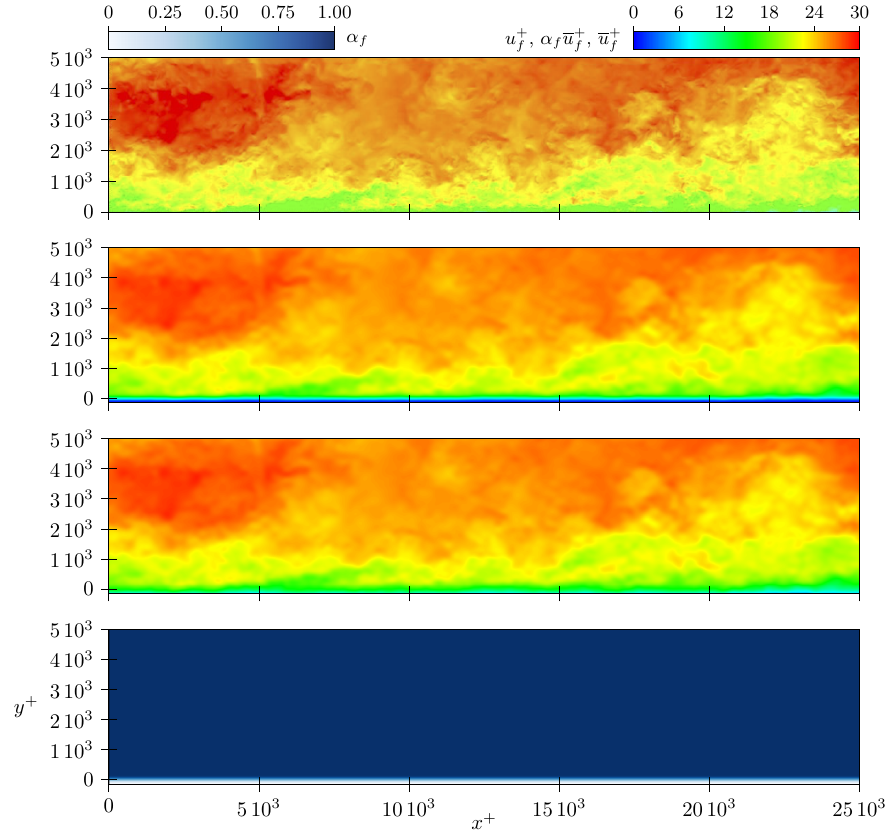}
  \caption{Instantaneous snapshots of the streamwise unfiltered, velocity $u^+_f(\boldsymbol{x})$, superficial fluid velocity $\alpha_f\overline{u}^+_f(\boldsymbol{x})$, filtered velocity $\overline{u}^+_f(\boldsymbol{x})$, and volume fraction $\alpha_f(\boldsymbol{x})$ for $\delta^+_f=300$ (top to bottom). Large scale structures remain in the filtered fields while fine scale structures are filtered out. $\overline{u}^+_f(\boldsymbol{x})$ field shows clearly that filtered quantities can be non-zero a distance of $\delta_f/2$ into the wall. $\alpha_f(\boldsymbol{x})$ field indicates fluid and solid regions with a transition region near the wall.}
  \label{fig:isocontours}
\end{figure}
We filter the DNS data using filter sizes from $\delta_f^+=50$ up to $300$ at regular intervals of $50$. To this end, we compute the volume-filtered quantities by direct application of the convolution (see (\ref{eq:def_filtered_I}) and (\ref{eq:def_filtered_u})) on a uniform grid such that $\Delta x=\Delta y=\Delta z=\delta_f/16$.  Note that ($\Delta x$,$\Delta y$,$\Delta z$) are the grid spacings in our post-processing, not to be confused with the DNS grid spacings ($\Delta x_\mathrm{DNS}$,$\Delta y_\mathrm{DNS}$,$\Delta z_\mathrm{DNS}$). We chose our post-processing resolution based on prior work in \citep{daveCharacterizationForcingSubfilter2025}, where we showed that numerical filtering errors become negligible with $\Delta x \leq\delta_f/16$. For $\delta_f^+\geq 200$, the post-processing grid spacing is coarser than the DNS grid spacings since $\Delta x^+=\delta_f^+/16=12.5$, whereas  $\Delta x^+_\mathrm{DNS}\sim 12$, $\Delta z^+_\mathrm{DNS}\simeq 6$, and $\Delta y^+_\mathrm{DNS}$ varies from $\sim 0.5$ to $\sim 10$.
For $\delta_f^+=50$ and $\delta_f^+=100$, the post-processing grid spacing fall below the DNS grid spacings, since $\Delta x^+=3.125$ and $\Delta x^+=6.25$ at $\delta_f^+=50$ and 100, respectively. In these cases, we use linear interpolations to evaluate the DNS data on the post-processing grid. However, to avoid contamination from interpolation errors we do not consider filter sizes smaller than $\delta_f^+=50$.
 
 Further, we extend the post-processing domain boundaries by a distance $\delta_f/2$ away from the top and bottom channel extents, and we place immersed solids in this extended region to represent the channel walls. The fluid-facing boundaries of these immersed solids are what we refer to as the ``wall'', or ``immersed boundary'' for the remainder of this manuscript. It is also important to note that in this \textit{a priori} analysis, the filtered quantities are computed at collocated points at cell centers as the DNS data available to us is also collocated at cell corners. In an \textit{a posteriori} analysis of the IBMLES framework, filtered quantities would be computed on a staggered grid.

We start by showing qualitatively the effect of volume-filtering using a coarse filter at $\delta_f^+=300$. Figure \ref{fig:isocontours} shows instantaneous snapshots of the unfiltered streamwise velocity $u^+_f=u_f/u_\tau$, the corresponding superficial and filtered streamwise velocities, $\alpha_f\overline{u}^+_f$ and $\overline{u}^+_f$ respectively, and volume fraction $\alpha_f$ in a wall-normal streamwise plane at an arbitrary time. The fluid volume fraction $\alpha_f$ clearly demarcates the fluid and solid regions where it is equal to 1 and 0, respectively. Close to the wall, there is a transition region of thickness $\delta_f^+=300$ resulting from volume-filtering the phase indicator function. At this coarse filter size, the superficial velocity $\alpha_f\overline{u}^+_f$ and filtered velocity $\overline{u}^+_f$ are devoid of finer scales present in the unfiltered field $u^+_f$. Yet, they retain many of the large scale structures seen in $u^+_f$. Close to the wall, $\alpha_f\overline{u}^+_f$ vanishes quickly, in part due to the decay of $\alpha_f$ premultiplying this quantity. Without $\alpha_f$ premultiplying $\overline{u}^+_f$, the value is significant at the wall ($y^+=0$), then decreases sharply to zero at position $y^+=-\delta_f^+/2$ into the solid. This illustrates the point that volume-filtered quantities may be significant up to a distance $\delta_f/2$ into the immersed solid.

\begin{figure}
    \centering
    \begin{subfigure}{0.495\linewidth}\centering
      \includegraphics[width=2.65in]{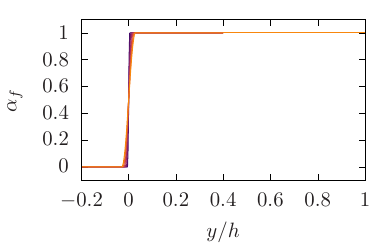}
      \caption{}
    \end{subfigure}
    \hfill
    \begin{subfigure}{0.495\linewidth}\centering
      \includegraphics[width=2.65in]{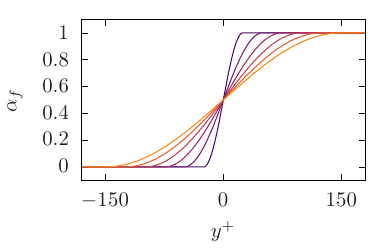}
      \caption{}
    \end{subfigure}
    \caption{Fluid volume fraction profiles for (a) channel half-height and (b) near-wall region at filter sizes $\delta_f^+=50$ ({\color{df050}\stl}), $\delta_f^+=100$ ({\color{df100}\stl}), $\delta_f^+=150$ ({\color{df150}\stl}), $\delta_f^+=200$ ({\color{df200}\stl}), $\delta_f^+=250$ ({\color{df250}\stl}), and $\delta_f^+=300$ ({\color{df300}\stl}). The volume fraction smoothly transitions from 0 to 1 over a distance of $\delta_f$ centered at the wall.}
    \label{fig:volume_fraction}
\end{figure}

The filter size controls the degree to which the wall region is diffused. Figure \ref{fig:volume_fraction} shows the effect of varying filter size on vertical profiles of fluid volume fraction $\alpha_f$ in the channel's lower half. Here, $\alpha_f=\alpha_f(y)$ varies with wall normal distance $y$ only, since the channel walls are flat and immobile. The transition from $\alpha_f=0$ to $\alpha_f=1$ occurs over a distance exactly equal to $\delta_f$ since the kernel has compact support. Since this flow is highly turbulent with $\Rey_\tau=5200$, the transition region remains narrow compared with the channel half-height even at the largest filter size $\delta_f^+=300$. At the walls ($y=0$ and $y=2h$), $\alpha_{f,w}=1/2$ in all cases, as half of the volume under the filter kernel is occupied by fluid regardless of the filter size. At positions $\delta_f/2\leq y\leq 2h-\delta_f/2$, $\alpha_f=1$ identically. Conversely, $\alpha_f=0$ at positions $y\leq -\delta_f/2$ and $y\geq 2h+\delta_f/2$. 

\subsection{Volume-filtered velocity statistics}

\begin{figure}
    \centering
    \begin{subfigure}{0.495\linewidth}\centering
      \includegraphics[width=2.65in]{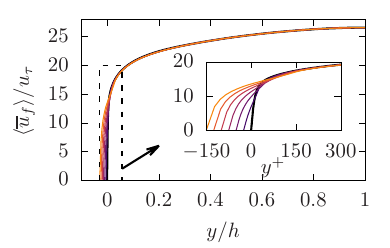}
      \caption{}
      \label{fig:u_mean_a}
    \end{subfigure}
    \hfill
    \begin{subfigure}{0.495\linewidth}\centering
      \includegraphics[width=2.65in]{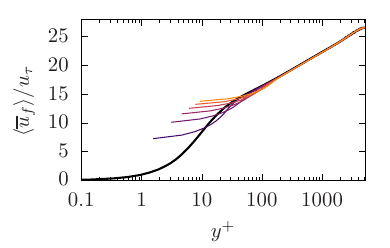}
      \caption{}
      \label{fig:u_mean_b}
    \end{subfigure}
    \caption{Profiles of normalized streamwise velocity in (a) linear and (b) semi-logarithmic scales at filter sizes $\delta_f^+=50$ ({\color{df050}\stl}), $\delta_f^+=100$ ({\color{df100}\stl}), $\delta_f^+=150$ ({\color{df150}\stl}), $\delta_f^+=200$ ({\color{df200}\stl}), $\delta_f^+=250$ ({\color{df250}\stl}), $\delta_f^+=300$ ({\color{df300}\stl}), and unfiltered DNS data ({\color{black}\stl}). The filtered velocity profiles converge to DNS profile away from the wall while significant slip is seen at the wall. All filtered profiles decrease to 0 at a distance $\delta_f/2$ into the wall.}
    \label{fig:u_mean}
\end{figure}

Next, we consider statistics of volume-filtered fluid velocity. To this end, we average all available flow snapshots from \citep{leeDirectNumericalSimulation2015} over the streamwise and spanwise directions, and between the channel's top and bottom halves. We use the notation $\langle \cdot \rangle$ to denote quantities averaged in this manner.

Figures \ref{fig:u_mean_a} and \ref{fig:u_mean_b} show $\langle \overline{u}_f^+\rangle$ profiles in linear and semi-logarithmic scales at filter sizes $\delta_f^+=50$ to 300, alongside $\langle u_f^+\rangle$ profile from DNS. From $y\sim \delta_f$ and further away from the wall, the mean volume-filtered velocity profiles collapse onto the DNS profile with all filters considered. The $\langle \overline{u}_f^+\rangle$ profiles also show a logarithmic region at distances $y^+> \delta_f^+$.

In contrast, we observe large differences close to the wall. As with the volume fraction field $\alpha_f$, $\langle \overline{u}_f^+\rangle$ extends beyond the wall and does not vanish until $y=-\delta_f/2$ into the solid. At the wall $y=0$, the mean slip velocity $\langle u_\mathrm{slip}\rangle =\langle \overline{u}_f\rangle|_{y=0} -\langle u_f\rangle|_{y=0}$ is significant, since $\langle u_\mathrm{slip}\rangle/u_{\tau}=O(10)$, and increases with $\delta_f$. From figure \ref{fig:u_mean_b}, $\langle \overline{u}_f\rangle$ profiles no longer show the viscous sublayer and buffer regions at these filter sizes. Capturing these regions requires filter sizes of the order of the DNS resolution, i.e., $\delta_f^+\sim 1$.

\begin{figure}
    \centering
    \begin{subfigure}{0.495\linewidth}\centering
      \includegraphics[width=2.65in]{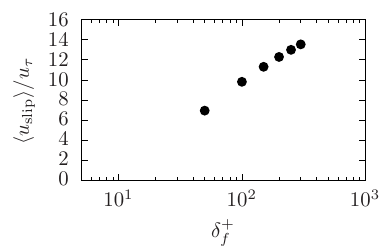}
      \caption{}
    \end{subfigure}
    \hfill
    \begin{subfigure}{0.495\linewidth}\centering
      \includegraphics[width=2.65in]{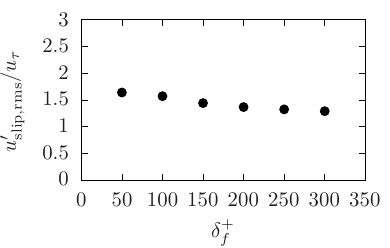}
      \caption{}
    \end{subfigure}
    \begin{subfigure}{0.495\linewidth}\centering
      \includegraphics[width=2.65in]{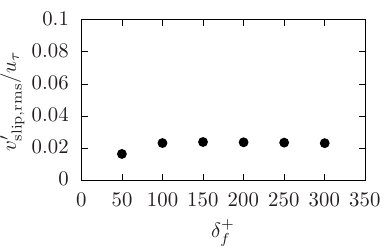}
      \caption{}
    \end{subfigure}
    \hfill
    \begin{subfigure}{0.495\linewidth}\centering
      \includegraphics[width=2.65in]{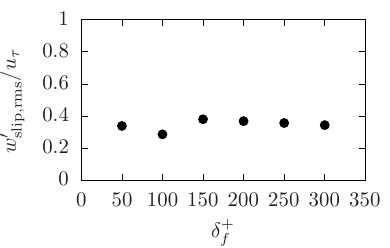}
      \caption{}
    \end{subfigure}
    \caption{Evolution of (a) the mean streamwise slip and (b) streamwise, (c) wall-normal, and (d) spanwise RMS slip fluctuations with filter size. Streamwise slip velocity is significant and shows logarithmic scaling at large $\delta_f^+$. Streamwise slip velocity fluctuations are approximately $10\%$ or less of mean slip values while spanwise and wall normal fluctuations are negligible.}
    \label{fig:u_slip}
\end{figure}

With filter sizes $\delta_f^+\geq 50$, it is clear that the no-slip condition is no longer valid for the volume-filtered velocity. To further clarify this point, we report the effect of varying filter size on the mean slip and root-mean-square (RMS) slip fluctuations in figure \ref{fig:u_slip}. Note that we only report the mean streamwise slip since we found the mean spanwise and wall normal slip to be negligible. At $\delta_f^+=50$, the wall-scaled steamwise slip velocity is $\langle u_\mathrm{slip}\rangle/u_{\tau}=6.93$. Increasing the filter size to $\delta_f^+=300$ increases the wall-scaled streamwise slip to $\langle u_\mathrm{slip}\rangle/u_{\tau}=13.5$. We also note that the streamwise slip shows a logarithmic trend with increasing $\delta_f^+$, as points in the logarithmic layer contribute greatly to the slip velocity at those larger filter sizes. 
Although much smaller than the mean, the streamwise slip fluctuations may be considered significant since $u'_{\mathrm{slip,rms}}/u_\tau\approx1.5$ for all filter sizes. However, wall-normal and spanwise slip fluctuations are negligibly small since $v'_{\mathrm{slip,rms}}/u_\tau\approx0.02$ and $w'_{\mathrm{slip,rms}}/u_\tau\approx 0.3$, respectively. Slip fluctuations appear to plateau for $\delta_f^+\geq 100$, which indicates that these fluctuations may be caused by flow structures much larger than the filter sizes we considered.

\begin{figure}
    \centering
    \includegraphics[width=5.2in]{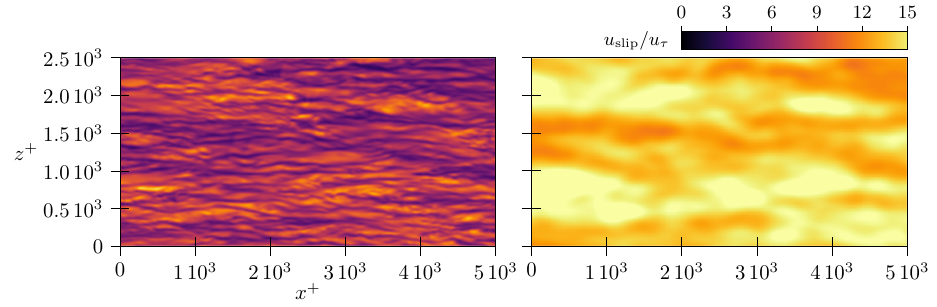}
    \caption{Instantaneous snapshots of slip velocity $u_\mathrm{slip}/u_{\tau}$ for $\delta_f^+=50$ (left) and $300$ (right). The slip velocity shows the imprint of inner wall coherent flow structures, such as high and low speed streaks at $\delta_f^+=50$, and the imprint of outer layer structures $\delta_f^+=300$.}
    \label{fig:u_slip_contour}
\end{figure}

Far from being uniform or random, the slip velocity represents an imprint of the flow structures on the wall for relatively small filter sizes. At large filter sizes, most of the near-wall structures are filtered out, and the slip velocity field is representative of outer layer flow structures. We illustrate this in figure \ref{fig:u_slip_contour} where we show instantaneous snapshots of slip velocity at $\delta_f^+=50$ and $300$ in a region of size $5000\times 2500$ in wall units. Clearly, the slip velocity field at $\delta_f^+=50$ is dominated by fine scale structures whereas the slip velocity field at $\delta_f^+=300$ displays only large scale structures. At $\delta_f^+=50$, the slip velocity field contains fine-scale structures that resemble high and low speed streaks characteristic of near-wall turbulence. In contrast, the slip velocity field at $\delta_f^+=300$ displays higher-speed large scale structures that represent the imprint of outer layer flow structures on the wall. The non-uniformities in this case are much lower than at $\delta_f^+=50$.

\begin{figure}
    \centering
    \begin{subfigure}{0.495\linewidth}\centering
      \includegraphics[width=2.65in]{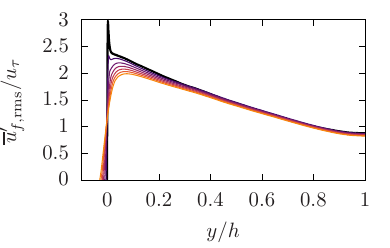}
      \caption{}
    \end{subfigure}
    \hfill
    \begin{subfigure}{0.495\linewidth}\centering
      \includegraphics[width=2.65in]{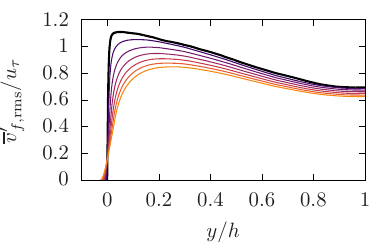}
      \caption{}
    \end{subfigure}
    \begin{subfigure}{0.495\linewidth}\centering
      \includegraphics[width=2.65in]{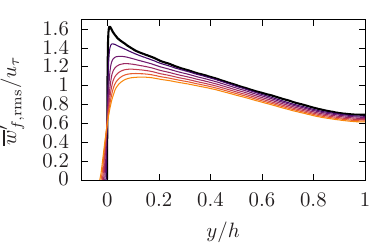}
      \caption{}
    \end{subfigure}
    \caption{Profiles of (a) streamwise, (b) wall normal, and (c) spanwise RMS filtered velocity fluctuations for different filter sizes. The filtered profiles show the same trends as DNS profiles with better agreement near the centerline. The fluctuations decrease with increasing filter size. Line colors as in figure \ref{fig:u_mean}.}
    \label{fig:rms_fluctuations}
\end{figure}

Figure \ref{fig:rms_fluctuations} shows the RMS velocity fluctuations from DNS and volume-filtered fields.
With increasing filter size, the RMS fluctuations in all three directions decrease since volume-filtering suppresses fluctuations that occur at length scales smaller than the filter size. The volume-filtered RMS profiles show the closest agreement with unfiltered profiles near the centerline. That is because fluctuations near the centerline are generated by large scale motions that are little affected by filtering at $\delta_f^+\leq 300$. The discrepancy between filtered and unfiltered RMS profiles is greatest close to the wall. In this region, near-wall fluctuations occur at smaller length scales. Filtering with $\delta_f^+=50$ up to 300 removes a growing proportion of these fluctuations, which leads to lower RMS values. All filtered profiles have non-zero values at the wall and decrease smoothly to zero at $\delta_f/2$ into the wall as volume under the filter no longer contains fluid. 

\subsection{Volume-filtered stress balance}

We now focus on analyzing terms that appear in the volume-filtered momentum equation (\ref{eq:filtered_momenutm}). Ensemble averaging this equation and considering the streamwise balance, we obtain the following balance,
\begin{eqnarray}
    \alpha_f\left(
    \mu_f\frac{d\left<\overline{u}_f\right>}{dy}
    +\left< R_{\mu,yx}\right>
    -\rho_f\left<\overline{v}'_f\overline{u}'_f\right>
    -\rho_f\left<\tau_{\mathrm{sfs},yx}\right>\right)
    =&&\nonumber\\
    && \hspace{-26ex}\tau_w\int_{-\infty}^y\left(-\frac{\alpha_f}{h}+\kernel_1(y')+\kernel_1(y'-2h)\right)dy'.
    \label{eq:momentum_balance}
\end{eqnarray}
We provide a detailed derivation in appendix \ref{sec:appendix_a}. The volume-filtered momentum balance (\ref{eq:momentum_balance}) contains contribution from the resolved viscous stress ($\mu_fd\left<\overline{u}_f\right>/dy$), residual viscous stress ($\left< R_{\mu,yx}\right>$), Reynolds stress ($-\rho_f\left<\overline{v}'_f\overline{u}'_f\right>$), and subfilter-scale stress ($-\rho_f\left<\mathsf{\tau}_{\mathrm{sfs},yx}\right>$). The sum of all 4 contributions, which we refer to as the total stress, is equal to a function that depends on the wall normal distance $y$, wall shear stress $\tau_w$, channel half-height $h$, as well as fluid volume fraction $\alpha_f$ and filter $\kernel_1$. The terms in $\kernel_1$ originate from the streamwise IB bodyforce $\left<F_{\mathrm{IB},x}\right>$. These terms are only active in a narrow band of width $\delta_f$ around the walls (at $y=0$ and $y=2h$), but play an important role in setting the correct profile in these regions. Away from walls, the total stress approaches the linear function observed in DNS, i.e.,
\begin{eqnarray}
    &&\tau_w\int_{-\infty}^y\left(-\frac{\alpha_f}{h}+\kernel_1(y')+\kernel_1(y'-2h)\right)dy'\approx
    \tau_w\left(1-\frac{y}{h}\right) \\
    &&\hspace{38ex} \quad (\text{for } \delta_f/2 \lesssim y \lesssim 2h-\delta_f/2)\nonumber
    \label{eq:linear_profile}
\end{eqnarray}
Close to walls, the total stress has a rounded profile that causes it to vanish inside the solid, i.e., at $y=-\delta_f/2$ and $y=2h+\delta_f/2$. 

\begin{figure}
    \centering
    \begin{subfigure}{0.495\linewidth}\centering
      \includegraphics[width=2.65in]{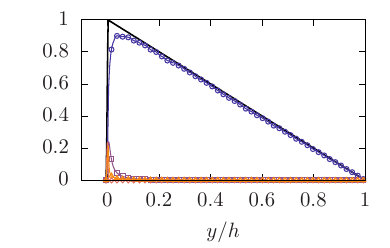}
      \caption{}
    \end{subfigure}
    \hfill
    \begin{subfigure}{0.495\linewidth}\centering
      \includegraphics[width=2.65in]{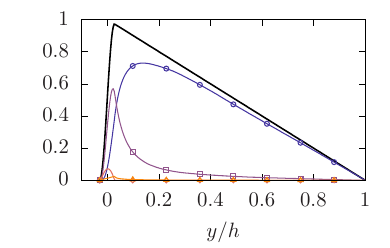}
      \caption{}
    \end{subfigure}
    \begin{subfigure}{0.495\linewidth}\centering
      \includegraphics[width=2.65in]{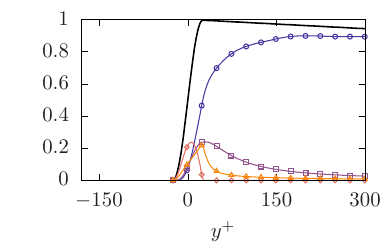}
      \caption{}
    \end{subfigure}
    \hfill
    \begin{subfigure}{0.495\linewidth}\centering
      \includegraphics[width=2.65in]{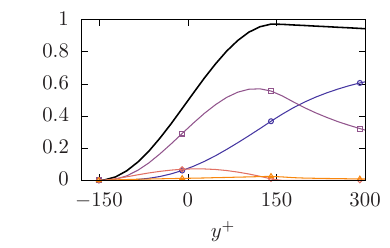}
      \caption{}
    \end{subfigure}
    \caption{Mean stress balance for $\delta^+_f=50$ (a,c) and $300$ (b,d) where (a,b) show profiles along the channel half-height and (c,d) show profiles in the near-wall region. The lines correspond to: normalized residual viscous stress $\alpha_f\left< R_{\mu,yx}\right>/\tau_w$ ({\color{var3}\stldiamonds}), normalized viscous stress $\alpha_f\mu_f(d\left<\overline{u}_f\right>/dy)/\tau_w$ ({\color{var4}\stltriangles}), normalized Reynolds shear stress $-\alpha_f\rho_f\left<\overline{v}'_f\overline{u}'_f\right>/\tau_w$ ({\color{var1}\stlcircles}), normalized subfilter-scale stress $-\alpha_f\rho_f\left<\mathsf{\tau}_{\mathrm{sfs},yx}\right>/\tau_w$ ({\color{var2}\stlsquares}), and total filtered stress normalized by $\tau_w$ ({\color{black}\stl}). Total stress shows a linear profile away from the wall with a transiton region that scales with $\delta_f$ near the wall. Increased filter size reduces contribution of viscous and Reynolds stresses and increases contribution of subfilter-scale and residual viscous stresses.}
    \label{fig:stress_balance}
\end{figure}

Figure \ref{fig:stress_balance} shows stress profiles at the finest ($\delta_f^+=50$) and coarsest ($\delta_f^+=300$) filter sizes tested. In both cases, the total stress has a narrow transition region at the wall and, otherwise, follows the linear profile from (\ref{eq:linear_profile}). As anticipated, this transition region is very narrow at $\delta_f^+=50$ and wider at $\delta_f^+=300$. At $\delta_f^+=50$, the filtered Reynolds stress is comparable to the unfiltered one, while the combination of subfilter-scale, filtered viscous and residual viscous stresses are comparable to unfiltered viscous stress. At $\delta_f^+=300$, the filtered Reynolds and viscous stresses decrease significantly which causes the subfilter-scale stress to adjust accordingly. At this larger filter size, the residual viscous stress is more significant than the filtered viscous stress although both have significantly reduced contribution compared to $\delta_f^+=50$.

\begin{figure}
    \centering
    \begin{subfigure}{0.495\linewidth}\centering
      \includegraphics[width=2.65in]{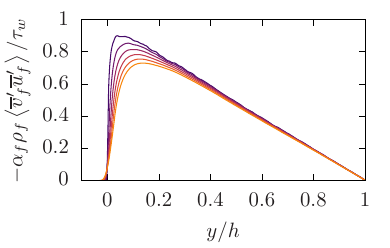}
      \caption{}
    \end{subfigure}
    \hfill
    \begin{subfigure}{0.495\linewidth}\centering
      \includegraphics[width=2.65in]{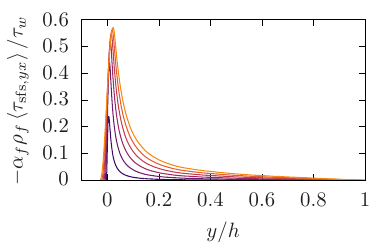}
      \caption{}
    \end{subfigure}
    \begin{subfigure}{0.495\linewidth}\centering
      \includegraphics[width=2.65in]{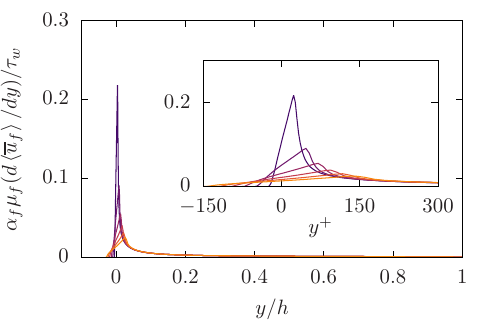}
      \caption{}
    \end{subfigure}%
    \hfill
    \begin{subfigure}{0.495\linewidth}\centering
      \includegraphics[width=2.65in]{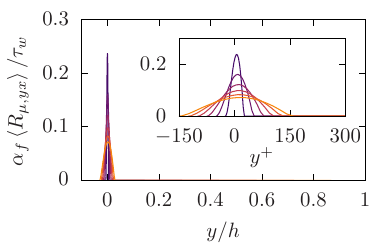}
      \caption{}
    \end{subfigure}
    \caption{Profiles of (a) Reynolds, (b) subfilter-scale, (c) filtered viscous, and (d) residual viscous stresses for filter sizes $\delta^+_f=50$ - $300$. Line colors as in figure \ref{fig:volume_fraction}.}
    \label{fig:momentum_contributions}
\end{figure}

Lastly, to analyze the effect of filter size on the different filtered stresses, we plot in figure \ref{fig:momentum_contributions} the filtered viscous, residual viscous, Reynolds and subfilter-scale stress profiles for all filter sizes tested. The filtered Reynolds stress has the shape expected from a classical channel flow where it dominates in the freestream but is negligible in the near-wall region relative to the other stress contributions. Increasing the filter size lowers filtered Reynolds stress and shifts its peak location towards the centerline. The subfilter-scale stress has a form that is reminiscent of the unfiltered viscous stress, where it is largest close to the wall and vanishes towards the centerline. Increasing the filter size causes this term to gain in magnitude as its normalized peak grows from 0.24 at $\delta_f^+=50$ to 0.57 at $\delta_f^+=300$. The filtered viscous stress and residual viscous stress are significant only near the wall. Although the latter is larger across all filter sizes tested, both terms have comparable normalized magnitudes. With increasing filter size, both residual and filtered viscous stress drop in magnitude. Note that while the filtered viscous stress has a long vanishing tail that extends towards the channel center, the residual viscous stress has compact support centered on the region from $-\delta_f/2$ to $\delta_f/2$ due to our choice of compact filter.

%% file: models.tex
\section{Modeling the wall-slip}\label{sec:models}

Modeling the wall-slip is in essence a turbulence wall model. As we show in \S\ref{sec:a_priori}, the streamwise slip velocity is significant with filter sizes $\delta_f^+\geq 50$. This makes modeling the wall-slip of paramount importance in volume-filtering LES.

In this section, we present two modeling strategies: (i) an algebraic slip velocity model and (ii) a slip-length model. 

%% file: models_algebraic.tex
\subsection{Algebraic slip velocity model}\label{sec:algebraic_model}

A simple model for the mean wall slip velocity $\left<u_\mathrm{slip}\right>$ can be formulated by volume-filtering an assumed velocity profile. In this study, we use the Van Driest model \citep{vandriestTurbulentFlowWall1956}, which gives the mean streamwise velocity profile as
\begin{eqnarray}
    \frac{\left<u^\mathrm{VD}\right>(y)}{u_\tau}=\int_0^{y/\delta_\nu}\frac{2}{1+\sqrt{1+4(\kappa y'(1-\exp(-y'/A^+)))^2}}dy',\label{eq:VD_profile}
\end{eqnarray}
where $\kappa = 0.41$, $A^+=26$, and $\delta_\nu=\nu/u_\tau$. Applying the volume-filter to this profile, we obtain the following slip model 
\begin{gather}
    \frac{\left<u_\mathrm{slip}^\mathrm{VD}\right>(\delta_f)}{u_\tau}=\frac{1}{\alpha_{f,w}}\int_0^{\delta_f/2}\int_0^{y/\delta_\nu}\frac{2}{1+\sqrt{1+4(\kappa y'(1-\exp(-y'/A^+)))^2}}\kernel_1(y)dy'dy. \label{eq:algebraic_model}
\end{gather}

\begin{figure}
    \centering
    \begin{subfigure}{0.495\linewidth}\centering
      \includegraphics[width=2.65in]{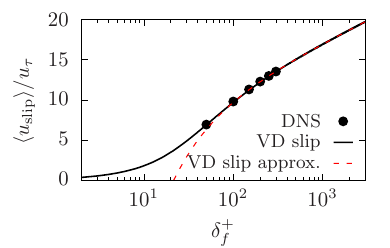}
      \caption{}
      \label{fig:VanDriestSlip}
    \end{subfigure}
    \hfill
    \begin{subfigure}{0.495\linewidth}\centering
      \includegraphics[width=2.65in]{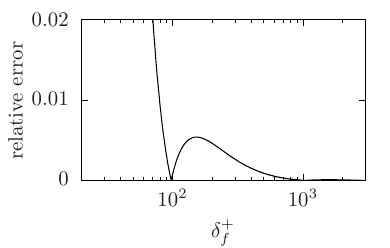}
      \caption{}
      \label{fig:VanDriestApprox}
    \end{subfigure}
    \caption{(a) Normalized streamwise slip velocity computed by filtering DNS data, numerical evaluation of the Van Driest slip (\ref{eq:algebraic_model}), and the Van Driest slip approximation (\ref{eq:approx_3}), and (b) relative error for the Van Driest slip approximation at different filter sizes. The Van Driest slip model shows excellent agreement with the filtered DNS data. The Van Driest approximation (\ref{eq:approx_3}) yields less than $1\%$ error for $\delta_f^+\geq100$ but diverges for $\delta_f^+\rightarrow 0$.}
    \label{fig:van_driest}
\end{figure}

The Van Driest slip model (\ref{eq:algebraic_model}) is a very good model for channel flow turbulence and similarly simple flow configurations. To show this, we report in figure \ref{fig:VanDriestSlip} comparison of the mean slip velocity obtained by directly volume-filtering the turbulent channel flow DNS $\left<u_\mathrm{slip}\right>$ and the slip derived from the Van Driest model $\left<u_\mathrm{slip}^\mathrm{VD}\right>$. Clearly, there is excellent agreement between the two for filter sizes $\delta_f^+=50$ to 300.

Despite reproducing the streamwise slip very well, the Van Driest slip model has a major inconvenience, which is that the nested double integrals in (\ref{eq:algebraic_model}) cannot be computed analytically. While numerical integration is acceptable in \textit{a priori} analyses, it is too costly and impractical in forward LES runs.

To alleviate this issue, we propose an approximation of (\ref{eq:algebraic_model}) that is much simpler to evaluate, yet provides accurate results for filter sizes $\delta_f^+\geq 50$. First, we note that for very large filter sizes, the Van Driest slip $\left<u_\mathrm{slip}^\mathrm{VD}\right>$ tends towards a logarithmic profile, i.e.,
\begin{equation}
  \left<u_\mathrm{slip}^\mathrm{VD}\right>/u_\tau \approx \frac{1}{\alpha_{f,w}}\frac{1}{2\kappa}\log(\delta_f^+/2) \quad \text{for}\;\delta_f^+\gg1. \label{eq:approx_1}
\end{equation}
Thus, we can express the Van Driest slip in terms of the logarithmic term above and a remainder function $f(\delta_f^+)$ such that
\begin{equation}
  \left<u_\mathrm{slip}^\mathrm{VD}\right>/u_\tau = \frac{1}{\alpha_{f,w}}\left(\frac{1}{2\kappa}\log(\delta_f^+/2) +f(\delta_f^+)\right). \label{eq:approx_2}
\end{equation}
We determine an approximation for $f$ using a power fit. Injecting it in (\ref{eq:approx_2}), we obtain the following approximation for the Van Driest slip
\begin{equation}
  \left<u_\mathrm{slip}^\mathrm{VD,approx}\right>/u_\tau = \frac{1}{\alpha_{f,w}}\left(\frac{1}{2\kappa}\log(\delta_f^+/2) + 0.972 - 72.0\left(1/\delta^+_f\right)^{0.9523} \right) \label{eq:approx_3}
\end{equation}
Figure \ref{fig:VanDriestApprox} shows the relative error between the Van Driest slip approximation $\left<u_\mathrm{slip}^\mathrm{VD,approx}\right>$ from (\ref{eq:approx_3}) and the exact Van Driest slip $\left<u_\mathrm{slip}^\mathrm{VD}\right>$ from (\ref{eq:algebraic_model}) is less than 1\% for filter sizes $\delta_f^+\geq 100$ but diverges for $\delta_f^+\rightarrow 0$.

%% file: models_slip_length.tex
\subsection{Slip-length model}\label{sec:slip_length_model}
This second strategy is a variation on the slip model proposed by \citet{boseDynamicSlipBoundary2014} and later expanded on by \citet{baeDynamicSlipWall2019}. \citet{boseDynamicSlipBoundary2014} suggested that a wall model can be formulated by finding a so-called slip-length $\ell_x$ such that the following relationship holds for the streamwise slip
\begin{eqnarray}
  \left< u_\mathrm{slip}\right>= \ell_x \left.\left<\frac{\partial \overline{u}_f}{\partial y}\right>\right|_w
\end{eqnarray}
They arrived at this conclusion by matching leading order terms in Taylor series expansions of the wall slip and filtered velocity wall gradient. A detailed description of a dynamic method of determining this slip-length is given in \citep{baeDynamicSlipWall2019}, and will not be discussed in this paper. Though this approach has been shown to yield good results in cases with simple walls, namely, turbulent channel flows,  the approach of \citet{boseDynamicSlipBoundary2014} suffers from commutation errors discussed in \S\ref{sec:framework}, and does not account for non-planar, moving, or deforming boundaries.

Here, we derive a slip-length model using different arguments from  \citet{boseDynamicSlipBoundary2014} and in a way that is adapted to the volume-filtering framework. Applying identity (\ref{eq:identity_grad_ver2}) with $\boldsymbol{\phi}=\boldsymbol{u}_f$ and evaluating at the wall, we get
\begin{eqnarray}
    \left.\overline{\nabla\boldsymbol{u}_f}\right|_w - \left.\nabla \overline{\boldsymbol{u}}_f\right|_w = \frac{1}{\alpha_{f,w}} \left(\left.\nabla\alpha_f\right|_w \right)\boldsymbol{u}_\mathrm{slip}.
\end{eqnarray}
It is noteworthy that the tensorial identity (\ref{eq:identity_grad_ver2}) is general since it applies to planar, non-planar, static, and moving walls alike, provided that the filter size $\delta_f$ is small enough to resolve the local curvature and any variation of the local wall velocity. Next, we take the dot product of the wall-normal unit vector with (\ref{eq:identity_grad_ver2}), taking care to multiply from the left. Rearranging the equation, we obtain a relationship for the slip velocity in the form
\begin{equation}
    \boldsymbol{u}_\mathrm{slip} = \frac{\alpha_{f,w}}{\left.\partial\alpha_f/\partial n\right|_w}\left(
    \left.\overline{\frac{\partial \boldsymbol{u}_f}{\partial n}}\right|_w
    -
    \left.\frac{\partial \overline{\boldsymbol{u}}_f}{\partial n}\right|_w
    \right). \label{eq:local_slip}
\end{equation}

To formulate our slip model, we introduce the non-dimensional coefficients $\boldsymbol{\lambda}=(\lambda_{n},\lambda_{t_1},\lambda_{t_2})$, such that
\begin{eqnarray}
    \lambda_{n} = \frac{\left<\left.\overline{\partial u_{f,n}/\partial n}\right|_w\right>}{\left<\left.\partial\overline{u}_{f,n}/\partial n\right|_w\right>};
    \quad
    \lambda_{t_1} = \frac{\left<\left.\overline{\partial u_{f,t_1}/\partial n}\right|_w\right>}{\left<\left.\partial\overline{u}_{f,t_1}/\partial n\right|_w\right>};
    \quad
    \lambda_{t_2} = \frac{\left<\left.\overline{\partial u_{f,t_2}/\partial n}\right|_w\right>}{\left<\left.\partial\overline{u}_{f,t_2}/\partial n\right|_w\right>},
    \label{eq:44}
\end{eqnarray}
where the `$n$' notation signifies components in the wall-normal direction, the `$t_1$' and `$t_2$' notations signify components in the wall-parallel directions. Substituting (\ref{eq:44}) into ensemble-averaged (\ref{eq:local_slip}) gives a slip formulation similar to the one in \citep{boseDynamicSlipBoundary2014}, i.e.,
\begin{eqnarray} 
    \left< u_{\mathrm{slip},n} \right>=\ell_{n} \left<\left.\frac{\partial\overline{u}_{f,n}}{\partial n}\right|_w\right>,\label{eq:slip_basic_1}\\ 
    \left< u_{\mathrm{slip},t_1} \right>=\ell_{t_1} \left<\left.\frac{\partial\overline{u}_{f,t_1}}{\partial n}\right|_w\right>,\label{eq:slip_basic_2}\\ 
    \left< u_{\mathrm{slip},t_2} \right>=\ell_{t_2} \left<\left.\frac{\partial\overline{u}_{f,t_2}}{\partial n}\right|_w\right>,\label{eq:slip_basic_3}    
\end{eqnarray}
and where the slip-lengths $\boldsymbol{\ell}=(\ell_{n},\ell_{t_1},\ell_{t_2})$ are
\begin{eqnarray} 
    \boldsymbol{\ell} &=& \frac{\alpha_{f,w}}{\left.\partial\alpha_f/\partial n\right|_w}(\boldsymbol{\lambda}-1).\label{eq:slip_basic_4}  
\end{eqnarray}
Non-dimensionalising by wall units, we obtain  the wall-scaled slip velocities and slip-lengths
\begin{eqnarray}
    \frac{\left< u_{\mathrm{slip},n} \right>}{u_\tau}&=&\ell_{n}^+ \left<\left.\frac{\partial\overline{u}_{f,n}^+}{\partial n^+}\right|_w\right>,\label{eq:slip_model_1_gen}\\
    \frac{\left< u_{\mathrm{slip},t_1} \right>}{u_\tau}&=&\ell_{t_1}^+ \left<\left.\frac{\partial\overline{u}_{f,t_1}^+}{\partial n^+}\right|_w\right>,\label{eq:slip_model_2_gen}\\
    \frac{\left< u_{\mathrm{slip},t_2} \right>}{u_\tau}&=&\ell_{t_2}^+ \left<\left.\frac{\partial\overline{u}_{f,t_2}^+}{\partial n^+}\right|_w\right>,\label{eq:slip_model_3_gen}\\
    \boldsymbol{\ell}^+ &=& \frac{\alpha_{f,w}}{\left.\partial\alpha_f/\partial n^+\right|_w}(\boldsymbol{\lambda}-1).  \label{eq:slip_model_4_gen}
\end{eqnarray}
Equations (\ref{eq:slip_model_1_gen})--(\ref{eq:slip_model_4_gen}) represent the wall-scaled slip velocities and slip-lengths in general wall coordinates for geometry that has slight curvature compared to the filter size. For the channel flow configuration presented here: the `$t_1$' direction corresponds to the streamwise direction $x$, the `$n$' direction corresponds to the wall-normal direction $y$, and the `$t_2$' direction corresponds to the spanwise direction $z$. Thus, the wall-scaled slip velocities and slip-lengths have the form 
\begin{eqnarray}
    \frac{\left< u_{\mathrm{slip}} \right>}{u_\tau}&=&\ell_x^+ \left<\left.\frac{\partial\overline{u}_f^+}{\partial y^+}\right|_w\right>,\label{eq:slip_model_1}\\
    \frac{\left< v_{\mathrm{slip}} \right>}{u_\tau}&=&\ell_y^+ \left<\left.\frac{\partial\overline{v}_f^+}{\partial y^+}\right|_w\right>,\label{eq:slip_model_2}\\
    \frac{\left< w_{\mathrm{slip}} \right>}{u_\tau}&=&\ell_z^+ \left<\left.\frac{\partial\overline{w}_f^+}{\partial y^+}\right|_w\right>,\label{eq:slip_model_3}\\
    \boldsymbol{\ell}^+ &=& \frac{\alpha_{f,w}}{\kernel_1^+(0)}(\boldsymbol{\lambda}-1),  \label{eq:slip_model_4}
\end{eqnarray}
where we also use the relationship $\left(\partial \alpha_f/\partial y^+\right)|_w=\kernel_1^+(0)$ for planar walls. Correspondingly, the non-dimensional coefficients $\boldsymbol{\lambda}=(\lambda_x,\lambda_y,\lambda_z)$ are
\begin{eqnarray}
    \lambda_x = \frac{\left<\left.\overline{\partial u_f/\partial y}\right|_w\right>}{\left<\left.\partial\overline{u}_f/\partial y\right|_w\right>};
    \quad
    \lambda_y = \frac{\left<\left.\overline{\partial v_f/\partial y}\right|_w\right>}{\left<\left.\partial\overline{v}_f/\partial y\right|_w\right>};
    \quad
    \lambda_z = \frac{\left<\left.\overline{\partial w_f/\partial y}\right|_w\right>}{\left<\left.\partial\overline{w}_f/\partial y\right|_w\right>}.
    \label{eq:44_ver2}
\end{eqnarray}

Equations (\ref{eq:slip_model_1})--(\ref{eq:slip_model_4}) relate the slip velocities to the wall normal gradients, volume fraction at the wall $\alpha_{f,w}$, wall-scaled filter size through $\kernel_1^+(0)\propto 1/\delta_f^+$, and non-dimensional coefficients $(\lambda_x,\lambda_y,\lambda_z)$. Among these dependencies, only $(\lambda_x,\lambda_y,\lambda_z)$ are unclosed. Therefore, closing these coefficients with an appropriately chosen model also closes the slip-lengths $(\ell^+_x,\ell^+_y,\ell^+_z)$.

It is clear that $\boldsymbol{\lambda}$ must depend on the filter size $\delta_f$, and possibly on the friction Reynolds number $\Rey_\tau$. Thus, we can write formally
\begin{equation}
	\boldsymbol{\lambda}=\boldsymbol{\lambda}(\delta_f^+,\Rey_\tau,\dots)
\end{equation}
In the limit of vanishingly small filter sizes,
\begin{equation}
 (\lambda_x,\lambda_y,\lambda_z) \rightarrow (1,1,1) \quad \text{for}\; \delta_f\rightarrow 0
\end{equation}
since both nominators and denominators in equations (\ref{eq:44_ver2}) become equal to one another. This ensures that the slip-lengths also vanish ($\boldsymbol{\ell}\rightarrow 0$) when $\delta_f\rightarrow 0$. Given that the data set we use in this study corresponds to a single $\Rey_\tau$, we are unable to tease out the exact functional dependence of $\boldsymbol{\lambda}$ on $\Rey_\tau$, but expect $\Rey_\tau$ dependence to be second order to $\delta_f^+$ dependence. There can also be additional dependencies in the case of non-planar, rough, moving, or deforming walls, although we do not investigate these effects in this study.

Although a universal model for $(\lambda_x,\lambda_y,\lambda_z)$ may be difficult to formulate, we propose next closures that apply to the turbulent channel flow and similarly simple wall bounded flows.

The \textit{a priori} analysis in \S\ref{sec:a_priori} shows that the wall-normal and spanwise slip velocities are negligible in a turbulent channel flow. Thus, the slip-lengths $\ell_y$ and $\ell_z$ are negligible. A model that reproduces this observation consists in simply taking  $\lambda_y$ and $\lambda_z$ constant and unitary i.e.,
\begin{equation}
  \lambda_y = 1 \quad\text{and}\quad \lambda_z = 1.
\end{equation}

\begin{figure}
    \centering
    \includegraphics[width=5.2in]{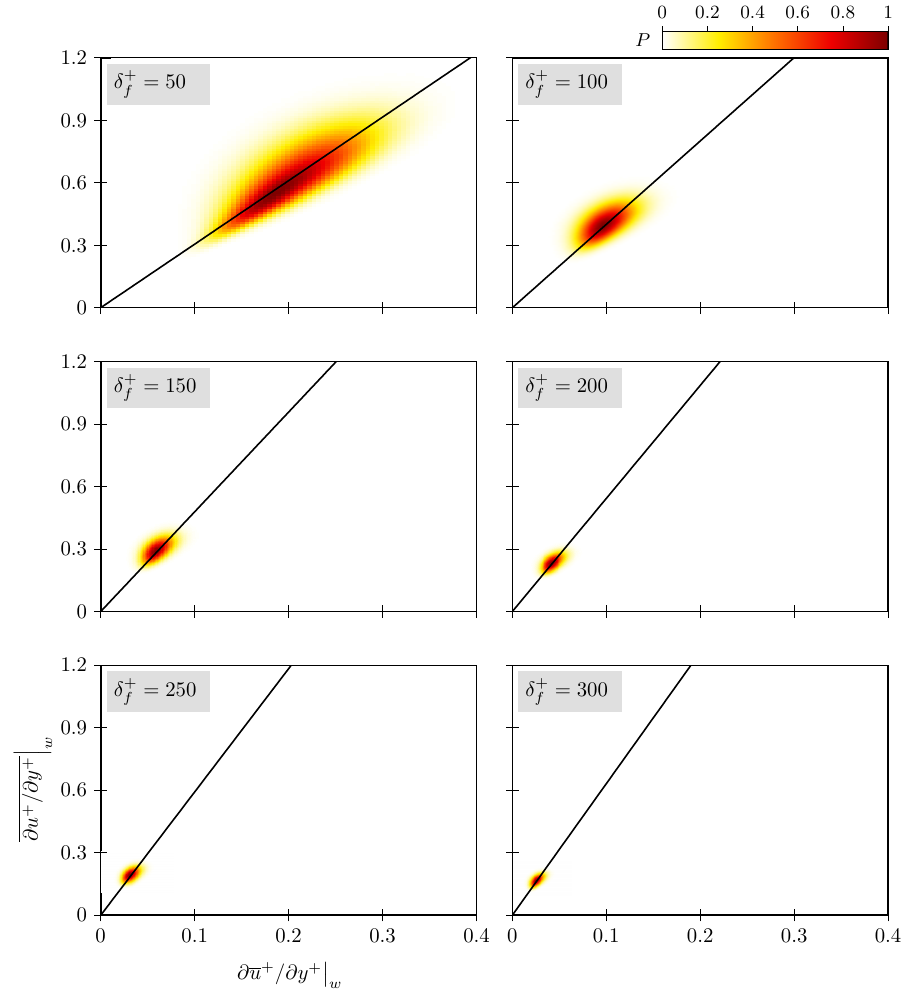}
    \caption{Joint PDFs of $\left.\overline{\partial u_f^+/\partial y^+}\right|_w$ and $\left.\partial \overline{u}_f^+/\partial y^+\right|_w$ normalized by peak values at $\delta_f^+=50-300$ (left to right, top to bottom). The oblong PDF distribution shows clear correlation between $\left.\overline{\partial u_f^+/\partial y^+}\right|_w$ and $\left.\partial \overline{u}_f^+/\partial y^+\right|_w$ at all filter sizes.}
    \label{fig:jpdfs}
\end{figure}

\begin{figure}
    \centering
    \begin{subfigure}{0.495\linewidth}\centering
      \includegraphics[width=2.65in]{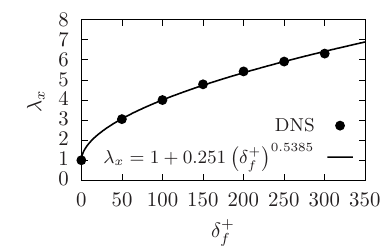}
      \caption{}
      \label{fig:lambda_ell_a}
    \end{subfigure}
    \hfill
    \begin{subfigure}{0.495\linewidth}\centering
      \includegraphics[width=2.65in]{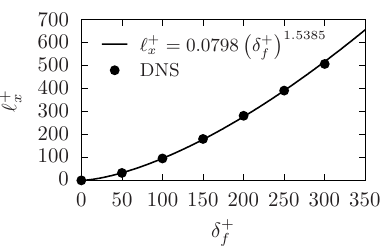}
      \caption{}
      \label{fig:lambda_ell_b}
    \end{subfigure}
    \caption{Variation of (a) the non-dimensional coefficient $\lambda$ and (b) normalized slip-length $\ell^+$ as a function of $\delta_f^+$ computed from filtered DNS data along with accompanying power law fits. Power law fits for $\lambda_x$ and $\ell^+_x$ show good agreement for all filter sizes tested with $\lambda_x$ scaling as $\left.\delta_f^+\right.^{0.5385}$ and $\ell^+_x$ scaling as $\left.\delta_f^+\right.^{1.5385}$ at large filter sizes.}
    \label{fig:lambda_ell}
\end{figure}

To elucidate the functional dependence of the streamwise $\lambda_x$ on $\delta_f^+$, we compute and report in figure \ref{fig:jpdfs} joint PDFs of $\left.\overline{\partial u_f^+/\partial y^+}\right|_w$ and $\left.\partial \overline{u}_f^+/\partial y^+\right|_w$. At $\delta_f^+=50$, the joint PDF isocontours are elongated and titled upwards, which shows a clear correlation between $\left.\overline{\partial u_f^+/\partial y^+}\right|_w$ and $\left.\partial \overline{u}_f^+/\partial y^+\right|_w$. The peak value is located at $\approx (0.2,0.6)$. With increasing filter size, the peak value shifts towards left and downwards, i.e., towards lower $\left.\overline{\partial u_f^+/\partial y^+}\right|_w$ and $\left.\partial \overline{u}_f^+/\partial y^+\right|_w$ values. This is because increasing filter size suppresses further the small scales responsible for large velocity gradients as explained in \S\ref{sec:a_priori}. The joint PDF iso-contours are also elongated at an angle that gets steeper with increasing filter size. 

From the joint PDFs, we extract the lines of best fit shown in black in figure \ref{fig:jpdfs}. The slope of these lines gives the non-dimensional coefficient $\lambda_x$.
From there, we determine $\ell^+_x$ with (\ref{eq:slip_model_4}) using $\alpha_{f,w}=1/2$, and $\kernel_1^+(0)=\pi/(2\delta_f^+)$ for the cosine filter. Figure \ref{fig:lambda_ell_a} shows the resulting $\lambda_x$ values for different filter sizes. As anticipated, $\lambda_x$ is equal to 1 at $\delta_f=0$ and increases with $\delta_f$. A power fit gives a good approximation of the functional dependence of $\lambda_x$ on $\delta_f$:
\begin{equation}
  \lambda_x = 1 + 0.251 \left.\delta_f^+\right.^{0.5385}
\end{equation}
Injecting this form in (\ref{eq:slip_model_4}) gives the following slip-length model
\begin{equation}
  \ell^+_x = 0.0798 \left.\delta_f^+\right.^{1.5385}
\end{equation}
for the cosine filter. Figure \ref{fig:lambda_ell_b} shows that this model captures indeed well the slip-length that we obtain directly by volume-filtering the DNS data.

%% file: filter_type.tex
\section{Effect of different filter kernels}\label{sec:effect_filter}
In this section, we investigate the impact of the filter kernel choice on slip velocity statistics, Van Driest slip model in \S\ref{sec:algebraic_model}, and slip-length model in \S\ref{sec:slip_length_model}. Here, we show that varying the filter kernel causes only small quantitative differences. The observations in \S\ref{sec:a_priori}, models, and scaling laws in \S\ref{sec:models} continue to hold.

Keeping with the same spirit as in \S\ref{sec:a_priori}, we consider filters that express as the product of 1D compact filters, i.e., following the decomposition (\ref{eq:filter_decomposition}). In addition to the cosine filter (\ref{eq:cosine_filter}), we consider the following ones, 
\begin{eqnarray}
   \text{Triangle:}\quad &&\kernel_1(r)=\left\{\begin{array}{@{}>{$}p{1.5in}<{$}r@{}} 
      \frac{2}{\delta_f}\left(1-2\left|r\right|/\delta_f\right) & |r|< \delta_f/2 \\
      0 & |r|\geq\delta_f/2 
   \end{array}\right.\\
   \text{Parabolic:}\quad &&\kernel_1(r)=\left\{\begin{array}{@{}>{$}p{1.5in}<{$}r@{}} 
      \frac{3/2}{\delta_f}\left(1-\left(2r/\delta_f\right)^2\right) & |r|< \delta_f/2 \\
      0 & |r|\geq\delta_f/2 
   \end{array}\right.\\
   \text{Triweight:}\quad &&\kernel_1(r)=\left\{\begin{array}{@{}>{$}p{1.5in}<{$}r@{}} 
       \frac{35/16}{\delta_f}\left(1-\left(2r/\delta_f\right)^2\right)^3 & |r|<\delta_f/2 \\
       0 & |r|\geq\delta_f/2
   \end{array}\right.
\end{eqnarray}

\begin{figure}\centering
	\includegraphics[width=4in]{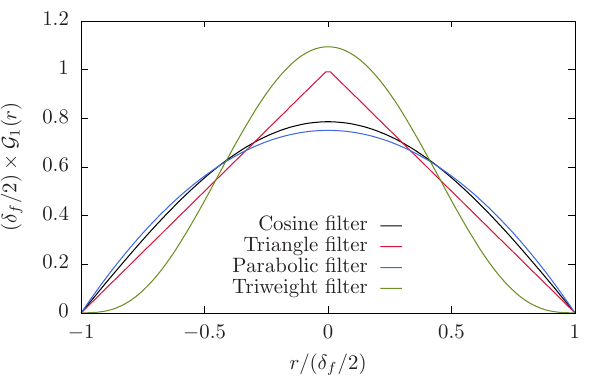}
	\caption{Filter kernels considered in this study. All filters are unitary, compact, and symmetric.}
	\label{fig:filter_types}
\end{figure}

Figure \ref{fig:filter_types} shows the morphology of these filter kernels. All kernels are symmetric and decrease monotonically from their peak value at the center to zero at $\pm\delta_f/2$. 
At first glance, the triangle kernel appears perhaps the simplest of all. It decreases linearly from center to edges and has a peak value $\kernel_1^\mathrm{triangle}(0)=2/\delta_f$. However, this kernel has a discontinuous derivative at the center $r=0$, which may complicate numerical evaluations of second and higher order derivatives of $\alpha_f$. The remaining filter kernels do not suffer from this issue. The cosine and parabolic kernels have similar shapes. Their peak values stand at $\kernel_1^\mathrm{parabolic}(0)= 1.5/\delta_f$ and $\kernel_1^\mathrm{cosine}(0)=(\pi/2)/\delta_f\approx 1.57/\delta_f$. The triweight kernel is the most concentrated at the center out of all kernels tested. It has a peak value of $\kernel_1^\mathrm{triweight}(0)=(35/16)/\delta_f\approx 2.19/\delta_f$ and decays faster at the edges than the other kernels.

\begin{figure}
    \centering
    \begin{subfigure}{0.495\linewidth}\centering
      \includegraphics[width=2.65in]{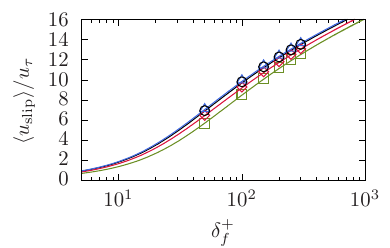}
      \caption{}
    \end{subfigure}
    \hfill
    \begin{subfigure}{0.495\linewidth}\centering
      \includegraphics[width=2.65in]{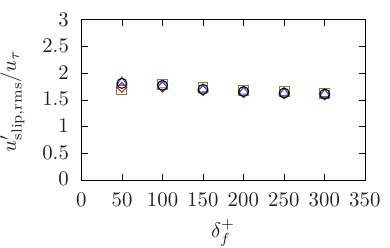}
      \caption{}
    \end{subfigure}
    \caption{Normalized (a) mean and (b) RMS fluctuations of the streamwise slip velocity for multiple filter kernels. Symbols correspond to filtered DNS data, while lines represent the Van-Driest model (\ref{eq:algebraic_model}). Colors correspond to different kernels as given in \ref{fig:filter_types}. Changing the filter kernel yields very little differences.}
    \label{fig:uSlip_comp}
\end{figure}

We show in figure \ref{fig:uSlip_comp} the slip velocity plotted with the associated Van-Driest model and RMS fluctuations for each filter kernel. Slip velocity trends are very similar for all filters, and they all show logarithmic scaling at large filter sizes. Profiles for cosine and parabolic filters are nearly identical. There are slightly lower values for the triangle and triweight filters with triweight having the lowest values overall. RMS fluctuations are nearly identical for all filter sizes. At $\delta_f^+=50$, fluctuations for triangle and triweight filters are slightly lower than those for cosine and parabolic filters similar to the overall trend seen in mean slip values.

\begin{figure}
    \centering
    \begin{subfigure}{0.495\linewidth}\centering
      \includegraphics[width=2.65in]{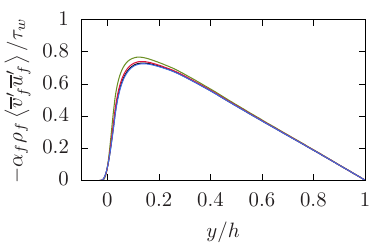}
      \caption{}
    \end{subfigure}
    \hfill
    \begin{subfigure}{0.495\linewidth}\centering
      \includegraphics[width=2.65in]{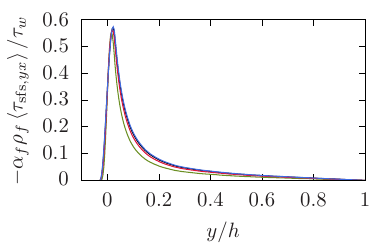}
      \caption{}
    \end{subfigure} \\
    \begin{subfigure}{0.495\linewidth}\centering
      \includegraphics[width=2.65in]{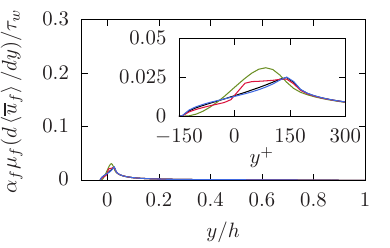}
      \caption{}
    \end{subfigure}
    \hfill
    \begin{subfigure}{0.495\linewidth}\centering
      \includegraphics[width=2.65in]{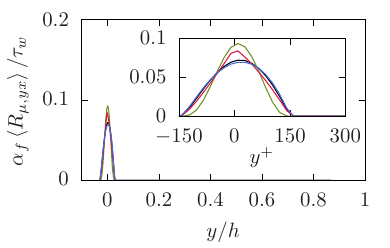}
      \caption{}
    \end{subfigure}
    \caption{Profiles of mean (a) Reynolds, (b) subfilter-scale, (c) filtered viscous, and (d) residual viscous stress (left to right, top to bottom) for $\delta^+_f=300$. Lines as in figure \ref{fig:filter_types}. Reynolds stress and subfilter-scale stress profiles are nearly identical for all filters tested. Filtered and residual viscous stresses show slight differences between filter kernels, although these differences remain less than 1\% in relative magnitude.}
    \label{fig:stress_comp}
\end{figure}

In addition to the slip velocity, we also investigate the effect of filter kernel choice on the terms in the momentum balance (\ref{eq:momentum_balance}), namely, the filtered viscous stress, the residual viscous stress, the Reynolds stress, and the subfilter-scale stress in figure \ref{fig:stress_comp}. Here too, the profiles change very little with different filter kernels. We provide only the data at $\delta_f^+=300$, since differences resulting from the kernel choice are most visible at the largest filter size. For the Reynolds and subfilter-scale stresses, the profiles are nearly identical. The only notable difference is the peak of the Reynolds stress with the triweight kernel is slightly higher than for other kernels, while the corresponding subfilter-scale stress is slightly lower. The residual and filtered viscous stress profiles are also quite similar for all filter kernels tested, although slight differences can be seen near the wall. Notably, the residual viscous stress profile has the same shape as the underlying filter kernel. The filtered viscous stress also varies a little with filter kernel. However, the magnitudes of these changes are very small as they are less than 1\% in all cases.

\begin{table}
  \centering
  \begin{tabularx}{\linewidth}{XXXX}
    Filter & $a_0$ & $b_0$ & $n_0$\\[1ex]
    Cosine & $0.972$ & $-72.0$ & $0.9523$ \\
    Triangle & $0.818$ & $-63.0$ & $0.8976$ \\
    Parabolic & $1.018$ & $-68.8$ & $0.9514$ \\
    Triweight & $0.607$ & $-70.1$ & $0.8930$ \\
  \end{tabularx}
  \caption{Table of coefficients in Van Driest slip approximation $\left<u_\mathrm{slip}^\mathrm{VD,approx}\right>/u_\tau = \frac{1}{\alpha_w}\left(\frac{1}{2\kappa}\log(\delta_f^+/2) + a_0 + b_0\left(1/\delta_f^+\right)^{n_0} \right)$ for cosine, triweight, triangle, and parabolic filters.}
  \label{table:2}
\end{table}

We show $\left<u_\mathrm{slip}^\mathrm{VD}\right>$ predicted by (\ref{eq:algebraic_model}) along with $\left<u_\mathrm{slip}\right>$ for each filter in figure \ref{fig:uSlip_comp}. There is excellent agreement between the two regardless of filter kernel. While the model is difficult to integrate, it provides accurate predictions so long as the filter used abides by the conditions in \S\ref{sec:framework}. Following the same procedure as for (\ref{eq:approx_3}), we find an approximation of the Van Driest slip velocity of the form
\begin{equation}
    \left<u_\mathrm{slip}^\mathrm{VD,approx}\right>/u_\tau = \frac{1}{\alpha_{f,w}}\left(\frac{1}{2\kappa}\log(\delta_f^+/2) + a_0 + b_0\left(1/\delta_f^+\right)^{n_0} \right) \quad \delta_f^+\geq 100
\end{equation}
where $a_0$, $b_0$, $n_0$ depend on the filter kernel chosen but vary little as one can see in table \ref{table:2}.  

\begin{figure}
    \centering
    \begin{subfigure}{0.495\linewidth}\centering
      \includegraphics[width=2.65in]{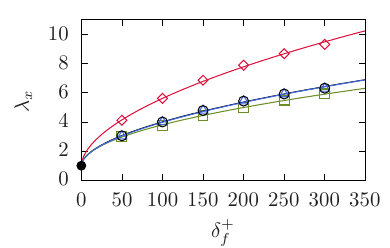}
      \caption{}
      \label{fig:model_comp_a}
    \end{subfigure}
    \hfill
    \begin{subfigure}{0.495\linewidth}\centering
      \includegraphics[width=2.65in]{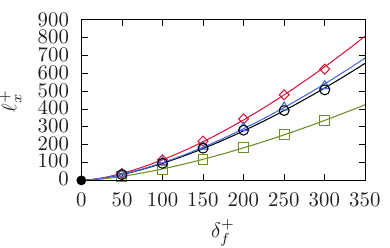}
      \caption{}
      \label{fig:model_comp_b}
    \end{subfigure}
    \caption{Variation of (a) $\lambda_x$ and (b) $\ell^+_x$ as a function of $\delta_f^+$ for different filter kernels. The same trends are obtained regardless of the kernel chosen. Colors as in figure \ref{fig:filter_types}. }
    \label{fig:model_comp}
\end{figure}

To assess the choice of filter type on the slip-length model, we plot in figure \ref{fig:model_comp} $\lambda_{x}$ and $\ell^+_{x}$ versus the normalized filter size $\delta_f^+$ for all filter kernels along with the associated power law fits. This leads to an expression of $\lambda_x$ of the form
\begin{equation}
    \lambda_x = 1 + a_1 {\delta_f^+}^{n_1},\label{eq:lambda_x_reln}
\end{equation}
where $a_1$ and $n_1$ depend on the filter kernel, which we provide in table \ref{table:1}. All fits show good agreement with results regardless of filter kernel. $\lambda_{x}$ curves are similar for all filter kernels. The triangle kernel yields the largest $\lambda_{x}$ values, likely due to its sharpness compared to the other kernels. Since $\ell^+_{x}$ derives directly from $\lambda_{x}$, the $\ell^+_{x}$ curves in figure \ref{fig:model_comp_b} also display similar trends for the cosine, parabolic, and triangle filters. Despite the differences one can see in figure \ref{fig:model_comp}, all $\lambda_{x}$ fits scale approximately as $1+0.3\left.\delta_f^+\right.^{0.5}$ and $\ell^+_{x}$ scales as approximately $0.08\left.\delta_f^+\right.^{1.5}$.

\begin{table}
  \centering
  \begin{tabularx}{\linewidth}{XXX}
    Filter & $a_1$ & $n_1$\\[1ex]
    Cosine & 0.251 & 0.539\\
    Triweight & 0.268 & 0.509 \\
    Triangle & 0.357 & 0.555\\
    Parabolic & 0.242 & 0.544
  \end{tabularx}
  \caption{Model parameters for the coefficient $\lambda=1+a_1{\delta_f^+}^{n_1}$ for different filter kernels.}
  \label{table:1}
\end{table}

%% file: dynamic_slip.tex
\section{Dynamic procedure for determining slip-length}\label{sec:dynamic}

To illustrate how the slip-length model may be used in IBMLES, we propose a dynamic procedure for determining slip-length that is derived in a procedure similar to that of \citet{boseDynamicSlipBoundary2014} and which leverages the model in (\ref{eq:lambda_x_reln}). We begin by introducing the test filtered stress tensor $\boldsymbol{T}_{\mathrm{sfs}}$ and test filtered SFS tensor $\widehat{\boldsymbol{\tau}}_{\mathrm{sfs}}$ in the following
\begin{eqnarray}
  \boldsymbol{T}_{\mathrm{sfs}}=\widehat{\overline{\boldsymbol{u}_f\boldsymbol{u}_f}}-\widehat{\overline{\boldsymbol{u}}}_f\widehat{\overline{\boldsymbol{u}}}_f,\label{eq:test_stress}\\
  \widehat{\boldsymbol{\tau}}_{\mathrm{sfs}}=\widehat{\overline{\boldsymbol{u}_f\boldsymbol{u}_f}}-\widehat{\overline{\boldsymbol{u}}_f\overline{\boldsymbol{u}}_f},\label{eq:test_sfs_stress}
\end{eqnarray}
where the `$\widehat{\phi}$' notation is used to denote a quantity $\phi$ that is test filtered with a filter size $\widehat{\delta}_f>\delta_f$. \textcolor{revision2}{Test filtering is performed in the same manner as (\ref{eq:def_filtered_u}), and it is assumed the nominal and test filter kernels, $\kernel_1$ and $\widehat{\kernel}_1$, are of the same type and differ only in filter size, $\delta_f$ and $\widehat{\delta}_f$.} Using the above definitions, we rearrange the Germano identity \citep{germanoDynamicSubgridscaleEddy1991} similar to what is done by \citet{boseDynamicSlipBoundary2014} to get
\begin{eqnarray}
  \widehat{\overline{\boldsymbol{u}_f\boldsymbol{u}_f}}=\widehat{\overline{\boldsymbol{u}}}_f\widehat{\overline{\boldsymbol{u}}}_f+\boldsymbol{T}_{\mathrm{sfs}}=\widehat{\boldsymbol{\tau}}_{\mathrm{sfs}}+\widehat{\overline{\boldsymbol{u}}_f\overline{\boldsymbol{u}}_f}.\label{eq:germano}
\end{eqnarray}
We then assume the slip relations (\ref{eq:slip_basic_1})--(\ref{eq:slip_basic_4}) apply instantaneously and are also valid for the test filtered level such that
\begin{eqnarray}
  \boldsymbol{u}_\mathrm{slip}=\mathrm{diag}(\boldsymbol{\ell})\left.\frac{\partial\overline{\boldsymbol{u}}_f}{\partial n}\right|_w;
  \quad
  \widehat{\boldsymbol{u}}_\mathrm{slip}=\mathrm{diag}(\widehat{\boldsymbol{\ell}})\left.\frac{\partial\widehat{\overline{\boldsymbol{u}}}_f}{\partial n}\right|_w,\label{eq:inst_slip_reln}\\
  \textcolor{revision2}{\boldsymbol{\ell}=\frac{\alpha_{f,w}}{\left.\partial\alpha_f/\partial n\right|_w}(\boldsymbol{\lambda}-1);}
  \quad
  \textcolor{revision2}{\widehat{\boldsymbol{\ell}}=\frac{\widehat{\alpha}_{f,w}}{\left.\partial\widehat{\alpha}_f/\partial n\right|_w}(\widehat{\boldsymbol{\lambda}}-1).}\label{eq:inst_slip_reln_2}  
\end{eqnarray}
where \textcolor{revision2}{$\widehat{\boldsymbol{u}}_\mathrm{slip}=\left.\widehat{\overline{\boldsymbol{u}}}_f\right|_w-\boldsymbol{u}_w$,} $\widehat{\boldsymbol{\ell}}$ \textcolor{revision2}{and $\widehat{\alpha}_f$ are the wall slip velocity,} slip-length, and \textcolor{revision2}{volume fraction, respectively,} at the test filtered level. From \S\ref{sec:slip_length_model}, the only assumption made about the wall up to this point is that it has slight curvature relative to the filter size. Based on the results of \S\ref{sec:a_priori}, $\ell_n=\widehat{\ell}_n=\ell_{t_2}=\widehat{\ell}_{t_2}=0$, so we only consider $\ell_{t_1}$ and $\widehat{\ell}_{t_1}$ which can be related to one another using (\ref{eq:lambda_x_reln}) as 
\begin{eqnarray}
  \widehat{\ell}_{t_1}=\frac{\widehat{\alpha}_{f,w}}{\alpha_{f,w}}\frac{\left.\partial\alpha_f/\partial n\right|_w}{\left.\partial\widehat{\alpha}_f/\partial n\right|_w}\left(\frac{\widehat{\delta}_f^+}{\delta_f^+}\right)^{n_1}\ell_{t_1},\label{eq:ell_t1_reln_gen}
\end{eqnarray}
where \textcolor{revision2}{we assume that (\ref{eq:lambda_x_reln}) applies at the test filtered level for $\widehat{\lambda}_{t_1}$}. For a planar wall, $\alpha_{f,w}=\widehat{\alpha}_{f,w}=1/2$, $\left.(\partial\alpha_f/\partial n)\right|_w=\kernel_1(0)\propto (1/\delta_f^+)$, and $\left.(\partial\widehat{\alpha}_f/\partial n)\right|_w=\widehat{\kernel}_1(0)\propto (1/\widehat{\delta}_f^+)$, where $\widehat{\kernel}_1$ represents the test filter kernel. Using these relations, (\ref{eq:ell_t1_reln_gen}) for a flat wall simplifies to 
\begin{eqnarray}
  \widehat{\ell}_{t_1}=\delta_R^{n_1+1}\ell_{t_1},\label{eq:ell_t1_reln}
\end{eqnarray}
where $\delta_R=\widehat{\delta}_f^+/\delta_f^+$. Evaluating (\ref{eq:germano}) at the wall and substituting (\ref{eq:inst_slip_reln}) for $\left.\widehat{\overline{u}}_{t_1}\right|_w=\widehat{u}_{\mathrm{slip,t_1}}+u_{w,t_1}$ yields
\begin{eqnarray}
  \left(\delta_R^{n_1+1}\ell_{t_1}\left.\frac{\partial\widehat{\overline{u}}_{t_1}}{\partial n}\right|_w+u_{w,t_1}\right)^2+\left.T_{\mathrm{sfs},t_1t_1}\right|_w-\left.\widehat{\tau}_{\mathrm{sfs},t_1t_1}\right|_w=\left.\widehat{\overline{u}_{t_1}\overline{u}_{t_1}}\right|_w.\label{eq:dyn_slip_length}
\end{eqnarray}
Solving (\ref{eq:dyn_slip_length}) for $\ell_{t_1}$ gives
\begin{eqnarray}
  \ell_{t_1}=\frac{-u_{w,t_1}+\sqrt{\left.\widehat{\overline{u}_{t_1}\overline{u}_{t_1}}\right|_w-\left.T_{\mathrm{sfs},t_1t_1}\right|_w+\left.\widehat{\tau}_{\mathrm{sfs},t_1t_1}\right|_w}}{\delta_R^{n_1+1}\left(\left.\partial\widehat{\overline{u}}_{t_1}/\partial n\right|_w\right)},\label{eq:dyn_ell_t1}
\end{eqnarray}
where we only consider the positive realizable root of $\ell_{t_1}$. For stationary walls, (\ref{eq:dyn_ell_t1}) can be further simplified with $u_{w,t_1}=0$.

While this model is derived in a similar nature to that of \citet{boseDynamicSlipBoundary2014}, it does not assume a linear dependence on the filter size ratio $\delta_R$ and instead relies on the $n_1$ scaling determined in \S\ref{sec:slip_length_model} and \S\ref{sec:effect_filter}. Another notable difference is that we do not assume the same slip-length for all coordinate directions, and (\ref{eq:dyn_slip_length}) can be solved explicitly instead of in a least-squares sense. We do not evaluate the dynamic procedure in this present analysis and save this for future \textit{a posteriori} analyses of IBMLES.  

%% file: conclusion.tex
\section{Conclusion}\label{sec:conclusion}

In this paper, we present a framework for LES with IBs to simulate flows over complex geometries at high Reynolds numbers. The elements that distinguish this strategy are the filter treatment near walls and modeling of the wall-slip velocity. The latter is essentially a model for the subfilter-scale near-wall flow and serves to impose the correct wall stress through the IB bodyforce. As such, we call this approach immersed boundary-modeled large eddy simulation (IBMLES).

Compared to other WRLES and WMLES methods, the present IBMLES is based on solving \emph{volume-filtered} transport equations. While volume-filtering is similar to traditional LES filtering in bulk regions of the flow, the treatment differs substantially close to walls. In particular, we truncate portions of the filter kernel that fall outside the fluid region, i.e, inside the solid, without changing its size or rescaling it. We also extend our definitions of filtered fluid quantities up to $\delta_f/2$ into the solid, since at those distances the filter kernel may still overlap with the fluid region.

Volume-filtering the Navier-Stokes equations and carefully accounting for all non-commutation terms, we are able to derive the IB bodyforce $\boldsymbol{F}_\mathrm{IB}$ without resorting to numerical heuristics. This also leads to the emergence of the SFS tensor $\bm{\mathsf{\tau}}_{\mathrm{sfs}}$ and an additional tensor, termed the residual viscous stress tensor $\mathsfbi{R}_\mu$. In this formulation, both $\boldsymbol{F}_\mathrm{IB}$ and $\bm{\mathsf{\tau}}_{\mathrm{sfs}}$ require closure. As we show in \S\ref{sec:ib_modeling}, modeling the wall-slip velocity $\boldsymbol{u}_\mathrm{slip}=\overline{\boldsymbol{u}}|_w -\boldsymbol{u}_w$ provides closure to $\boldsymbol{F}_\mathrm{IB}$.

To understand the nature of the wall-slip velocity, the effect of the residual viscous stress, and their significance at filter sizes $\delta_f^+=O(100)$, we perform an \textit{a priori} analysis using DNS data of turbulent channel flow at $\Rey_\tau=5200$ \citep{leeDirectNumericalSimulation2015}. Below, we summarize the key findings:
\begin{enumerate}
  \item The streamwise wall-slip velocity is significant at all filter sizes tested ($\delta_f^+=50$ -- 300). This indicates clearly that a model for the streamwise wall-slip velocity must be supplied to obtain a successful IBMLES. In the turbulent channel flow, the wall-normal and spanwise slip velocities are negligible. However, these components could be significant in other configurations such as those with curved or moving walls.
  \item The wall-slip velocity is far from being uniform, albeit the non-uniformities have a small relative magnitude. At smaller filter sizes ($\delta_f^+\approx 50$), the non-uniformities capture the imprint of inner layer flow structures on the wall. At larger filter size ($\delta_f^+\approx 300$), the signature of the outer layer flow structures supplants that of the inner layer flow structures.
  \item The SFS stresses in IBMLES are analogous to SFS stresses in traditional LES. We expect existing SFS models, such as the Dynamic Smagorinsky Model \citep{smagorinskyGeneralCirculationExperiments1963,germanoDynamicSubgridscaleEddy1991}, to apply equally well in IBMLES.
  \item The residual viscous stress $\mathsfbi{R}_\mu$ is non-zero only in a band of size $\delta_f$ centered on the wall. Nevertheless, it has a magnitude that is comparable or greater than that of the filtered viscous stress. Since it is fully closed and its calculation through equation (\ref{eq:residual_viscous_stress}) does not present significant challenge, it ought to be taken into account in IBMLES.
\end{enumerate}

The IB bodyforce $\boldsymbol{F}_\mathrm{IB}$ needs closure in IBMLES since it depends on the wall slip velocity (see equations (\ref{eq:forcing_ver1})--(\ref{eq:wall_stress})) and the latter is significant for filter sizes $\delta_f^+=O(100)$. The Van Driest and slip-length models we introduce in section \S\ref{sec:models} provide closure to the wall slip velocity, and hence the IB bodyforce. 

A good approximation of the streamwise slip velocity based on the \citet{vandriestTurbulentFlowWall1956} profile is
\begin{equation}
  \left<u_\mathrm{slip}^\mathrm{VD,approx}\right>/u_\tau \approx \frac{1}{\alpha_{f,w}}\left(\frac{1}{2\kappa}\log(\delta_f^+/2) + a_0 + b_0\left(1/\delta^+_f\right)^{n_0} \right) \text{ for } \delta_f^+\geq 100\label{eq:approx_conclusion}
\end{equation}
where $a_0=0.81\pm 0.21$, $b_0=-67.5\pm 4.5$, and $n_0=0.92\pm 0.03$ are constants of this model. Their values vary a little depending on the chosen filter kernel. In \textit{a priori} tests, this model captures very well the streamwise slip velocity in the turbulent channel flow at $\Rey_\tau=5200$. A drawback of this approach is that the approximation (\ref{eq:approx_conclusion}) is only valid for $\delta_f^+\geq 100$. Specifically, the slip in (\ref{eq:approx_conclusion}) diverges as $\delta_f^+\rightarrow 0$ instead of vanishing.

With the slip-length model, the mean slip velocities express as
\begin{equation}
  \left<\boldsymbol{u}_{\mathrm{slip}}\right> = \frac{\alpha_{f,w}}{\kernel_1(0)}\mathrm{diag}(\boldsymbol{\lambda}-1)\left<\left.\frac{\partial \boldsymbol{\overline{u}}_f}{\partial y}\right|_w\right>\label{eq:slip_length_conclusion}
\end{equation}
where $\mathrm{diag}(\boldsymbol{a})$ denotes the diagonal matrix with components $\boldsymbol{a}$. This approach is an extension of an earlier slip-length model proposed by \citet{boseDynamicSlipBoundary2014}. Here, the slip-lengths $\boldsymbol{\ell}=\alpha_{f,w}(\boldsymbol{\lambda}-1)/\kernel_1(0)=(\ell_x,\ell_y,\ell_z)$ are directly related to the volume fraction at the wall $\alpha_{f,w}$, which is equal to $1/2$ for perfectly flat walls, the peak filter kernel value $\kernel_1(0)\propto (1/\delta_f)$, and the non-dimensional coefficients $\boldsymbol{\lambda}=(\lambda_x,\lambda_y,\lambda_z)$.
Based on the \textit{a priori} tests in the turbulent channel flow, we find the following closures
\begin{eqnarray*}
  \lambda_x = 1 + a_1 {\delta_f^+}^{n_1};\quad 
  \lambda_y = 1;\quad 
  \lambda_z = 1,
\end{eqnarray*}
with $a_{1}=0.30\pm 0.06$ and $n_{1}=0.53\pm0.02$ depending on the filter kernel. This approach has the advantage of being valid for filter sizes $\delta_f^+=O(100)$ as well as in the limit $\delta_f^+\rightarrow 0$, where it reverts to a no-slip model. Thus, IBMLES with the slip-length closure above recovers the framework for DNS with IBs we introduced in \citep{daveVolumefilteringImmersedBoundary2023} in the limit $\delta_f^+\rightarrow 0$.

From the above closures, we find $\ell_y=\ell_z=0$, and the dynamic procedure for determining $\ell_x$ has the form
\begin{eqnarray}
  \ell_x=\frac{-u_{w}+\sqrt{\left.\widehat{\overline{u}\;\overline{u}}\right|_w-\left.T_{\mathrm{sfs},xx}\right|_w+\left.\widehat{\tau}_{\mathrm{sfs},xx}\right|_w}}{\delta_R^{n_1+1}\left(\left.\partial\widehat{\overline{u}}/\partial n\right|_w\right)},
\end{eqnarray}
which can be solved explicitly. While this procedure is not assessed in the present analysis, it has the advantage of determining slip values locally instead of providing a single mean value.  The performance and stability of this procedure will be assessed in future \textit{a posteriori} studies. 

It should be mentioned that both Van Driest and slip-length models provide only a single value for the mean slip velocity, despite the non-uniformity and unsteadiness of the local slip velocity. However, as we show in \S\ref{sec:a_priori}, these fluctuations are very small relative to the mean. Thus, for the purpose of modeling the wall slip velocity, the local slip may be suitably approximated by its mean using either the Van Driest or slip-length models. \textit{A posteriori} testing is required to confirm or deny this hypothesis as the IB bodyforce contains instantaneous terms along with the slip velocity (see equations (\ref{eq:forcing_ver1})--(\ref{eq:wall_stress})), which may require a slip model that captures the slight fluctuations to enforce the correct wall shear stress. The dynamic slip-length procedure outlined in \S\ref{sec:dynamic} may capture these slip fluctuations, although we do not assess this in the present study. 

It is also worthwhile to note the analysis presented is carried out using a single canonical dataset of turbulent channel flow at a single $\Rey_\tau$. For the slip-length model, the functional dependence of $\boldsymbol{\lambda}$ on $\Rey_\tau$ cannot be determined \emph{a priori} without considering additional datasets at larger $\Rey_\tau$ values. Nevertheless, we expect this dependence to be secondary to the $\delta_f^+$ dependence. For the Van-Driest model, the  $\Rey_\tau$ dependence is ``built-in'' through the Van Driest profile (see \S\ref{sec:algebraic_model}). 

Further, the planar turbulent channel flow, while convenient for model derivation, omits many of the physics found in more complicated flows, such as separated flows and flows over moving walls. Still, we expect both Van Driest and slip-length models to hold well, when the walls display small curvature or slow motion that does not lead to separation. In such cases, the filter must be chosen such that the local curvature is well resolved. Additionally, one would use the wall coordinate form of the slip-length model (see (\ref{eq:slip_model_1_gen})--(\ref{eq:slip_model_4_gen})) with the same scaling laws for the non-dimensional coefficients $\boldsymbol{\lambda}$ and determine $\left(\partial \alpha_f/\partial n^+\right)|_w$ from the $\alpha_f$ field. \textcolor{revision2}{For walls with significant curvature, one could refine the filter width to resolve the curvature, and we would expect the models to hold, but this may become prohibitively expensive for complex wall geometries. As these models do not include dependencies for rough, moving, or deforming walls, they likely will not hold for applications where these effects are significant. Additionally, it is possible that the model parameters $a_1$ and $n_1$ for the slip-length model could exhibit a dependence on the Reynolds number.}

With the theoretical formulation and associated closure models of IBMLES now established, our future efforts will focus on \textit{a posteriori} testing to quantify the extent to which the method can reproduce the dynamics of high–Reynolds number flows over complex and moving surfaces as well as the performance of the proposed dynamic slip procedure.

Lastly, during the preparation of this manuscript and following publication of our original work \citep{daveVolumefilteringImmersedBoundary2023} on the volume-filtering immersed boundary method, related ideas were considered by \citet{hausmannPhysicallyConsistentImmersed2024} and \citet{hausmannWallmodeledLargeEddy2025}. We emphasize, however, that these studies adopt substantially different treatments of the convective and viscous terms, requiring non-standard subfilter-scale stress tensors and additional modifications arising from lack of Galilean invariance in the governing equations they employ.

%% file: appendix.tex
\label{sec:appendix_a}
In this section, we present the derivation of the volume-filtered momentum balance in equation (\ref{eq:momentum_balance}).

First, note that the mean wall normal and spanwise momentum in a turbulent channel flow is zero. To get the streamwise balance, we apply ensemble averaging to the streamwise projection of the volume-filtered momentum equation (\ref{eq:filtered_momentum_ver1}). The left-hand side of this balance is
\begin{gather}
    \mathrm{LHS}=\left<\rho_f\left(\frac{\partial}{\partial t}(\alpha_f\overline{u}_f)+\frac{\partial}{\partial x}(\alpha_f\overline{u}_f\overline{u}_f)+\frac{\partial}{\partial y}(\alpha_f\overline{v}_f\overline{u}_f)+
    \frac{\partial}{\partial z}(\alpha_f\overline{w}_f\overline{u}_f)\right)\right>.
\end{gather}
The channel flow configuration analyzed is steady, fully-developed, and periodic in both the $x$ and $z$ directions. Additionally, $\alpha_f$ depends only on $y$ in this configuration ($\left<\alpha_f\right>=\alpha_f(y)$). Consequently, the LHS simplifies to
\begin{gather}
    \mathrm{LHS}=\frac{d}{dy}\left(\alpha_f\rho_f\left<\overline{v}_f\overline{u}_f\right>\right).
\end{gather}
Introducing fluctuations with respect to ensemble averaging as
\begin{eqnarray}
  \overline{u}'_f &=& \overline{u}_f - \left<\overline{u}_f\right>,\\
  \overline{v}'_f &=& \overline{v}_f - \left<\overline{v}_f\right>,
\end{eqnarray}
the left-hand side is then
\begin{gather}
    \mathrm{LHS}=\frac{d}{dy}\left(\alpha_f\rho_f\left<\overline{v}'_f\overline{u}'_f\right>\right).\label{eq:A2}
\end{gather}
The right-hand side of the streamwise momentum balance is
\begin{gather}
\mathrm{RHS}=
\boldsymbol{e}_x\cdot\left<\nabla\cdot\left(\alpha_f\left(\overline{\bm{\mathsf{\tau}}}^R_f+\mathsfbi{R}_{\mu}-\rho_f\bm{\mathsf{\tau}}_\mathrm{sfs}\right)\right)-\iint_{\boldsymbol{y}\in S_w}\boldsymbol{n}\cdot\bm{\mathsf{\tau}}_f(\boldsymbol{y},t)\kernel(\boldsymbol{x}-\boldsymbol{y})dS\right>,\label{eq:A3}
\end{gather}
where $\boldsymbol{e}_x$ is a unitary vector in the streamwise direction.
Following the same reasoning used previously, the first term simplifies to
\begin{eqnarray}
  \boldsymbol{e}_x\cdot\left<\nabla\cdot\left(\alpha_f\left(\overline{\bm{\mathsf{\tau}}}^R_f+\mathsfbi{R}_{\mu}-\rho_f\bm{\mathsf{\tau}}_\mathrm{sfs}\right)\right)\right>
  =&&\nonumber\\
   &&\hspace{-30ex}
  -\alpha_f\left<\frac{\partial\overline{p}}{\partial x}\right>
  +\frac{d}{dy}\left(
  \alpha_f\mu_f\frac{d\left<\overline{u}_f\right>}{dy}
  +\alpha_f\left< R_{\mu,yx}\right>
  -\alpha_f\rho_f\left<\tau_{\mathrm{sfs},yx}\right>
  \right),\label{eq:A4}
\end{eqnarray}
Further, the following identity holds in a turbulent channel flow
\begin{equation}
  \alpha_f\left<\frac{\partial\overline{p}}{\partial x}\right>
  =
  \alpha_f\left<\overline{\frac{\partial p}{\partial x}}\right>
  =-\alpha_f \frac{\tau_w}{h}\label{eq:A5}
\end{equation}

Since we are deriving this balance for a channel flow, we can split the total wall surface ($S_w$) into bottom ($S_\mathrm{bottom}$) and top walls ($S_\mathrm{top}$) and write the last term in (\ref{eq:A3}) as
\begin{eqnarray}
\boldsymbol{e}_x\cdot\left<\iint_{\boldsymbol{y}\in S_w}\boldsymbol{n}\cdot\bm{\mathsf{\tau}}_f(\boldsymbol{y},t)\kernel(\boldsymbol{x}-\boldsymbol{y})dS\right>=&&\nonumber\\
&&\hspace{-36ex}
\iint_{\boldsymbol{y}\in S_\text{bottom}}\left<\tau_{f,yx}\right>|_{bw}\kernel(\boldsymbol{x}-\boldsymbol{y})dS
-
\iint_{\boldsymbol{y}\in S_\text{top}}\left<\tau_{f,yx}\right>|_{tw}\kernel(\boldsymbol{x}-\boldsymbol{y})dS
\end{eqnarray}
Here, $\left<\tau_{f,yx}\right>|_{bw}$ and $\left<\tau_{f,yx}\right>|_{tw}$ denote the average shear stress on the bottom and top walls, respectively, and are equal to $\left<\tau_{f,yx}\right>|_{bw}=\tau_w$ and $\left<\tau_{f,yx}\right>|_{tw}=-\tau_w$. Introducing the kernel decomposition (\ref{eq:filter_decomposition}), we arrive at
\begin{eqnarray}
\boldsymbol{e}_x\cdot\left<\iint_{\boldsymbol{y}\in S_w}\boldsymbol{n}\cdot\bm{\mathsf{\tau}}_f(\boldsymbol{y},t)\kernel(\boldsymbol{x}-\boldsymbol{y})dS\right>=
\tau_w \left(\kernel_1(y) + \kernel_1(y-2h)\right)
\end{eqnarray}
Combining equations (\ref{eq:A2}), (\ref{eq:A4}), and (\ref{eq:A5}), we obtain the momentum balance as
\begin{eqnarray}
  \frac{d}{dy}\left(
  \alpha_f\mu_f\frac{d\left<\overline{u}_f\right>}{dy}
  +\alpha_f\left< R_{\mu,yx}\right>
  -\alpha_f\rho_f\left<\overline{v}'_f\overline{u}'_f\right>
  -\alpha_f\rho_f\left<\tau_{\mathrm{sfs},yx}\right>
  \right) &&\nonumber\\
  &&\hspace{-30ex}=\tau_w\left(
  -\frac{\alpha_f}{h}
  + \kernel_1(y) + \kernel_1(y-2h)\right).
\end{eqnarray}
Integrating this last equation gives the balance in equation (\ref{eq:momentum_balance}).